\documentclass[a4paper,12pt]{article}
\pdfoutput=1
\usepackage{url}
\usepackage{epsfig}
\usepackage{amsmath}
\usepackage{amssymb}
\usepackage{amsfonts}
\usepackage{bbm}
\usepackage[footnotesize]{caption}
\usepackage{graphicx}
\usepackage{mathrsfs}
\usepackage[font=scriptsize]{subcaption}
\usepackage{color}
\usepackage{comment}
\usepackage{mathtools}
\usepackage{braket}
\usepackage{cite} 
\usepackage{hyperref}
\usepackage{multirow}
\usepackage{relsize} 
\usepackage{fullpage}
\usepackage{makecell}

\numberwithin{equation}{section}
\newcommand{\secheadmath}[1]{\texorpdfstring{$#1$}{TEXT}} 
\newcommand{\overbar}[1]{\mkern 1.5mu\overline{\mkern-1.5mu#1\mkern-1.5mu}\mkern 1.5mu}
\newcommand{\pmatr}[1]{\begin{pmatrix} #1 \end{pmatrix}}

\newcommand{\xiphase}{\rho_{\xi}}
\newcommand{\relphase}{\rho_\mathrm{atm}\!-\!\rho_\mathrm{sol}}
\newcommand{\physphase}{\eta}
\newcommand{\xivev}{v_{\xi}}

\allowdisplaybreaks[2]
\begin{document}
\begin{titlepage}
\vspace*{0.7cm}

\begin{center}
{\bf\Large  Towards a complete ${\bf A_4\times SU(5)}$ SUSY GUT} \\[12mm]
Fredrik~Bj\"{o}rkeroth$^{\star}$\footnote{E-mail: {\tt f.bjorkeroth@soton.ac.uk}},
Francisco~J.~de~Anda$^{\dagger}$
\footnote{E-mail: \texttt{franciscojosedea@gmail.com}},
Ivo~de~Medeiros~Varzielas$^{\star}$
\footnote{E-mail: \texttt{ivo.de@soton.ac.uk}},
Stephen~F.~King$^{\star}$
\footnote{E-mail: \texttt{king@soton.ac.uk}}
\\[-2mm]

\end{center}
\vspace*{0.50cm}
\centerline{$^{\star}$ \it
School of Physics and Astronomy, University of Southampton,}
\centerline{\it
SO17 1BJ Southampton, United Kingdom }
\vspace*{0.2cm}
\centerline{$^{\dagger}$ \it
Departamento de F{\'i}sica, CUCEI, Universidad de Guadalajara, M{\'e}xico}
\vspace*{1.20cm}

\begin{abstract}
{\noindent
We propose a renormalisable model based on $A_4$ family symmetry with an
$SU(5)$ grand unified theory (GUT) which leads to the minimal supersymmetric standard model (MSSM) with a two right-handed neutrino seesaw mechanism.
Discrete $\mathbb{Z}_9\times \mathbb{Z}_6$ symmetry provides the fermion mass hierarchy
in both the quark and lepton sectors, while $\mathbb{Z}_4^R$ symmetry 
is broken to $\mathbb{Z}_2^R$, identified as usual R-parity. Proton decay is highly suppressed by these symmetries.
The strong CP problem is solved in a similar way to the Nelson-Barr mechanism.
We discuss both the $A_4$ and $SU(5)$ symmetry breaking sectors, including doublet-triplet splitting,
Higgs mixing and the origin of the $\mu$ term. The model provides an excellent fit (better than one sigma) to all quark and lepton (including neutrino) masses and mixing with spontaneous CP violation. 
With the $A_4$ vacuum alignments, $(0,1,1)$ and $(1,3,1)$, the model predicts the entire PMNS mixing matrix 
with no free parameters, up to a relative phase, selected to be $2\pi/3$ from a choice of the nine complex roots of unity,
which is identified as the leptogenesis phase. The model predicts a normal neutrino mass hierarchy with 
leptonic angles $\theta^l_{13}\approx 8.7^{\circ}$, $\theta^l_{12}\approx 34^{\circ}$, $\theta^l_{23}\approx 46^{\circ}$
and an oscillation phase $\delta^l\approx -87^{\circ}$.
} 
\end{abstract}
\end{titlepage}

\thispagestyle{empty}
\vfill
\newpage

\setcounter{page}{1}
\section{Introduction}
The Standard Model (SM), although highly successful, leaves many unanswered questions in its wake such as: what (if anything) stabilises the recently discovered Higgs boson mass?  
Are the three known gauge forces unified into a simple gauge group
which also explains charge quantisation? 
What is the origin of the three families of quarks and leptons and their pattern of masses, mixing and CP violation? 
Why is CP so accurately conserved by the strong interactions?
The answers to such questions may help resolve longstanding cosmological puzzles such as the nature of dark matter and the origin of matter-antimatter asymmetry, both of which are unexplained within the SM.

In this paper we propose a realistic and fairly complete model capable of addressing all the above questions unanswered by the SM. The basic ingredients of our model are Supersymmetry (SUSY) together with an $SU(5)$ Grand Unified Theory (GUT), flavoured by an $A_4$ family symmetry (for a review see e.g. \cite{King:2013eh}). The model is minimal in the sense that $SU(5)$ is the smallest GUT group and $A_4$ is the smallest family symmetry group that admits triplet representations. Also, below the GUT scale, the model yields the minimal supersymmetric standard model (MSSM) supplemented by a minimal two right-handed neutrino seesaw mechanism. The model is realistic in the sense that it provides a successful (and natural) description of the fermion mass and mixing spectrum, including spontaneous CP violation, while resolving the strong CP problem. It is fairly complete in the sense that GUT and flavour symmetry breaking are addressed, including doublet-triplet splitting, Higgs mixing and the origin of the MSSM $\mu$ term.  

The model also allows a WIMP dark matter candidate due to the conserved MSSM R-parity, and permits matter-antimatter asymmetry via leptogenesis involving the two right-handed neutrinos. We shall show that the leptogenesis phase is equal to the single phase appearing in the neutrino mass matrix, providing a direct link between neutrino oscillations and matter-antimatter asymmetry, although we shall not discuss cosmological aspects any further in this paper.

We emphasise the predictive nature of the model in the lepton sector, where the entire PMNS matrix is predicted without any free parameters, up to a discrete choice of a single phase. Large lepton mixing is accounted for by the seesaw mechanism \cite{Minkowski:1977sc} with constrained sequential dominance (CSD) \cite{King:1998jw,King:2005bj,Antusch:2011ic}). With a diagonal two right-handed neutrino mass matrix $M_R$, the dominant right-handed neutrino $\nu_R^{\rm atm}$ mainly responsible for the atmospheric neutrino mass $m_3$ has couplings to $(\nu_e, \nu_{\mu}, \nu_{\tau})$ proportional to $(0,1,1)$, while the subdominant right-handed neutrino $\nu_R^{\rm sol}$ giving the solar neutrino mass $m_2$ has couplings to $(\nu_e, \nu_{\mu}, \nu_{\tau})$ proportional to $(1,3,1)$. These couplings, corresponding to the so called CSD3 scheme \cite{King:2013iva,Bjorkeroth:2014vha}, originate from $A_4$ vacuum alignment.%
\footnote{CSD4 models have been discussed in \cite{King:2013hoa}}
The model consequently predicts a normal neutrino mass hierarchy,  $m_3>m_2\gg m_1=0$.

As mentioned above, the lepton sector is controlled by a relative phase which is selected to be $2\pi/3$, chosen from the nine complex roots of unity arising from spontaneous CP violation of a $\mathbb{Z}_9 \times \mathbb{Z}_{6}$ discrete symmetry, by a mechanism proposed in \cite{Antusch:2011sx}. Such a spontaneous CP violating scenario had been proposed previously in order to account for the smallness of CP violation in the soft SUSY sector \cite{Ross:2004qn}. We also employ a $\mathbb{Z}_4^R$ discrete R-symmetry (as the origin of MSSM R-parity, as in \cite{Lee:2011dya}) and a missing partner (MP) mechanism \cite{Masiero:1982fe} for doublet-triplet splitting as recently advocated for flavoured GUTs in \cite{Antusch:2014poa}. The model predicts very sparse lepton and down-type quark Yukawa matrices, with five texture zeroes, and Yukawa elements involving simple $SU(5)$ Clebsch-Gordan (CG) ratios of $4/9$ and $9/2$ for the first and second families, with $m_{\tau}/m_b=1$ for the third family, all in excellent agreement with their experimental values run up to the GUT scale \cite{Antusch:2009gu}. 

Quark mixing originates predominantly from a non-diagonal and naturally hierarchical up-type Yukawa matrix, provided by the broken $\mathbb{Z}_9$ discrete family symmetry. Quark CP violation, however, comes exclusively from a single off-diagonal element in the down Yukawa matrix. By contrast, to excellent approximation, all lepton mixing and CP violation originates from the neutrino mass matrix, whose structure is also controlled by the $A_4$ family symmetry and the $\mathbb{Z}_{6} $ symmetry via the CSD3 type vacuum alignment as described above \cite{King:2013iva}.

Although there have been many attempts in the literature based on $A_4$ flavoured $SU(5)$ SUSY GUTs (for an incomplete list see e.g. \cite{deMedeirosVarzielas:2005qg}), we would argue that none are as successful or complete as the present one. For example, many of the previous models predicted mixing very close to tri-bimaximal and are by now excluded. Indeed the present model is the first one based on CSD3 capable of predicting all the lepton mixing parameters consistent with current data on lepton mixing \cite{King:2013iva,Bjorkeroth:2014vha} (see also \cite{King:2013hoa}). The full literature on flavoured SUSY GUTs, i.e. which involve
a (discrete) family symmetry, is quite extensive (for an incomplete list see e.g. \cite{huge}). The goal of all these models is clear: to address the questions left unanswered by the SM. It will take some time and (experimental) effort to resolve all these models. However the most promising models are those that make testable predictions while being theoretically complete and consistent. 

While there are many different chiral superfields in this model, indeed almost exactly a hundred, it is important to note that we are explicitly presenting a renormalisable model. Any ``non-renormalisable terms'' generated below the Planck scale are required to have a specific well defined realization through multiple renormalisable terms involving heavy messenger fields that can be integrated out around the GUT scale. The respective effective theory after they are integrated out is actually more predictive than otherwise, with a normal neutrino mass hierarchy, a zero lightest neutrino mass, and all lepton mixing angles and CP phases predicted. The model presented here is amongst the most predictive and complete SUSY GUTs of flavour, consistent with current data.

The layout of the remainder of the paper is as follows: in Section \ref{sec:SM} we describe the superfields directly related to the SM fermions and neutrinos, as well as their Yukawa structures as imposed by the GUT and family symmetries when certain $A_4$ breaking vacuum expectation values (VEVs) are applied; we also perform a global fit to the parameters of the model and present our predictions for the lepton sector. In Section \ref{sec:A4} we describe the superfields that are responsible for breaking the family symmetries and how the required $A_4$ breaking VEVs arise. In Section \ref{sec:GUT} several aspects related to the GUT are discussed, particularly how to break $SU(5)$ and $\mathbb{Z}_4^R$ down to the MSSM with R-parity in a viable way (i.e. addressing doublet-triplet splitting, the origin of the $\mu$ term and proton decay). We also discuss the resolution to the strong CP problem. In Section \ref{sec:link} we discuss the link between leptogenesis and the oscillation phase in this model. Finally in Section \ref{sec:con} we summarise our main results and conclude. Appendix~\ref{A4} summarises the $A_4$ conventions used in this paper, in the basis of \cite{Ma:2001dn}.

\section{The Yukawa sector of the model \label{sec:SM}}
\begingroup

\newcommand{\five}{$\textrm{5}$}
\newcommand{\fivebar}{$\bar{\textrm{5}}$}
\newcommand{\ten}{10}
\newcommand{\fortyfive}{45}
\newcommand{\fortyfivebar}{\overbar{45}}
\newcommand{\twentyfour}{24}
\newcommand{\twentyfourbar}{\overbar{24}}
\newcommand{\onep}{1$^{\prime}$}
\newcommand{\onepp}{1$^{\prime\prime}$}

\begin{table}
\centering
\footnotesize
\begin{minipage}[b]{0.45\textwidth}
\centering
\captionsetup{width=\textwidth}
\begin{tabular}{| c | c c | c | c | c |}
\hline
\multirow{2}{*}{\rule{0pt}{4ex}Field}	& \multicolumn{5}{c |}{Representation} \\
\cline{2-6}
\rule{0pt}{3ex}			& $A_4$ & SU(5) & $\mathbb{Z}_9$ & $\mathbb{Z}_6$ & $\mathbb{Z}_4^R$ \\ [0.75ex]
\hline \hline
\rule{0pt}{3ex}%
$F$ 			& 3 & \fivebar 		& 0 & 0 & 1 \\
$T_1$ 			& 1 & 10		& 5 & 0 & 1 \\
$T_2$ 			& 1 & 10		& 7 & 0 & 1 \\
$T_3$ 			& 1 & 10		& 0 & 0 & 1 \\
$N_1^c$ 		& 1 & 1	 		& 7 & 3 & 1 \\
$N_2^c$ 		& 1 & 1 		& 8 & 3 & 1 \\
$\Gamma$		& 1 & 1			& 0 & 3 & 1 \\
\rule{0pt}{3ex}%
$H_5$			& 1 & \five		& 0 & 0 & 0 \\
$H_{\bar{5}}$		& 1 & \fivebar 		& 2 & 0 & 0 \\
$H_{\twentyfour}$ 	& \onep & 24	& 3 & 0 & 0 \\
$\Lambda_{24}$ 	& \onep 	& 24		& 0 & 0 & 0 \\
$H_{\fortyfive}$	 	& 1 & 45 	& 4 & 0 & 2 \\
$H_{\fortyfivebar}$ 	& 1 & $\fortyfivebar$ 	& 5 & 0 & 0 \\
\rule{0pt}{3ex}%
$\xi$ 			& 1 & 1			& 2 & 0 & 0 \\
$\theta_1$		& 1 & 1			& 1 & 3 & 0 \\
$\theta_2$		& 1 & 1			& 1 & 4 & 0 \\
\rule{0pt}{3ex}%
$\phi_e$		& 3 & 1 		& 0 & 0 & 0 \\
$\phi_\mu$		& 3 & 1		& 3 & 0 & 0 \\
$\phi_\tau$		& 3 & 1 		& 7 & 0 & 0 \\
$\phi_1$		& 3 & 1 		& 3 & 2 & 0 \\
$\phi_2$		& 3 & 1 		& 1 & 3 & 0 \\
$\phi_3$		& 3 & 1 		& 3 & 1 & 0 \\
$\phi_4$		& 3 & 1 		& 2 & 1 & 0 \\
$\phi_5$		& 3 & 1 		& 6 & 2 & 0 \\
$\phi_6$		& 3 & 1 		& 5 & 2 & 0 \\
\hline
\end{tabular}
\caption{Superfields containing quarks and leptons and symmetry breaking scalars.\\}
\label{ta:SMF}
\end{minipage}%
\qquad\begin{minipage}[b]{0.45\textwidth}
\centering
\captionsetup{width=\textwidth}
\begin{tabular}{| c | c c | c | c | c |}
\hline
\multirow{2}{*}{\rule{0pt}{4ex}Field}	& \multicolumn{5}{c |}{Representation} \\
\cline{2-6}
\rule{0pt}{3ex}			& $A_4$ & SU(5) & $\mathbb{Z}_9$ & $\mathbb{Z}_6$ & $\mathbb{Z}_4^R$ \\ [0.75ex]
\hline \hline
\rule{0pt}{3ex}%
$X_{1}$			& 1 		& \fivebar 	& 7 & 0 & 1 \\
$X_{2}$			& 1 		& \five 		& 2 & 0 & 1 \\
$X_{3}$ 		& 1 		& \fivebar	& 6 & 0 & 1 \\
$X_{4}$			& 1 		& \five		& 3 & 0 & 1 \\
$X_{5}$			& \onepp	& \fivebar 	& 3 & 0 & 1 \\
$X_{6}$			& \onep 	& \five 		& 6 & 0 & 1 \\
$X_{7}$			& 1 		& \fivebar	& 2 & 0 & 1 \\
$X_{8}$			& \onepp	& \five		& 7 & 0 & 1 \\
$X_{9}$			& \onep 	& \fivebar	& 0 & 0 & 1 \\
$X_{10}$		& \onep 	& \five		& 0 & 0 & 1 \\
\rule{0pt}{3ex}%
$X_{11}$		& 1 		& \fivebar	& 1 & 3 & 1 \\
$X_{12}$		& 1 		& \five		& 7 & 5 & 1 \\
$X_{13}$		& 1 		& \fivebar	& 2 & 3 & 1 \\
$X_{14}$		& 1 		& \five		& 6 & 5 & 1 \\
\rule{0pt}{3ex}%
$\Sigma_{1}$		& 1 		& \fivebar 	& 7 & 0 & 2 \\
$\Sigma_{2}$		& 1 		& \five		& 2 & 0 & 0 \\
$\Sigma_{3}$		& 1 		& \fivebar 	& 5 & 0 & 2 \\
$\Sigma_{4}$		& 1 		& \five 		& 4 & 0 & 0 \\
$\Sigma_{5}$		& 1 		& \fivebar	& 3 & 0 & 2 \\
$\Sigma_{6}$		& 1 		& \five		& 6 & 0 & 0 \\
$\Sigma_{7}$		& 1 		& \fivebar 	& 1 & 0 & 2 \\
$\Sigma_{8}$		& 1 		& \five 		& 8 & 0 & 0 \\
$\Sigma_{9}$		& 1 		& \fivebar 	& 8 & 0 & 2 \\
$\Sigma_{10}$		& 1 		& \five 		& 1 & 0 & 0 \\
$\Sigma_{11}$		& 1 		& \fivebar 	& 6 & 0 & 2 \\
$\Sigma_{12}$		& 1 		& \five 		& 3 & 0 & 0 \\
$\Sigma_{13}$		& 1 		& \fivebar 	& 4 & 0 & 2 \\
$\Sigma_{14}$		& 1 		& \five 		& 5 & 0 & 0 \\
$\Sigma_{15}$		& 1 		& \fivebar 	& 2 & 0 & 2 \\
$\Sigma_{16}$		& 1 		& \five 		& 7 & 0 & 0 \\
\hline	
\end{tabular}
\caption{Superfield messengers for the quark and lepton Yukawa couplings (and other GUT breaking couplings discussed in Section \ref{sec:splitting}).}
\label{ta:Mess}
\end{minipage}
\end{table}

\endgroup

The model involves an $A_4\times SU(5)$ CP invariant superpotential at the GUT scale, 
where all symmetries, including CP, are spontaneously broken along supersymmetric flat directions,
as discussed in Sections \ref{sec:A4} and \ref{sec:GUT}. 
As already noted, it involves a further 
$\mathbb{Z}_9 \times \mathbb{Z}_{6}$ discrete family symmetry as well as 
a $\mathbb{Z}_4^R$ discrete R-symmetry.
The purpose of this section is to describe those aspects of the model pertaining to the Yukawa sector,
i.e. relevant for understanding quark and lepton masses, mixing and CP violation.
The flavour sector of the model is very important in our approach, since we 
make a serious attempt to understand and, where possible, predict the experimentally observable
fermion masses and mixing matrices.

In Table \ref{ta:SMF} we show the matter superfields $F$, $T_i$ that contain the quarks and leptons,
as well as the right-handed neutrino superfields $N_i^c$ and double seesaw superfield
$\Gamma$, all of which carry unit $\mathbb{Z}_4^R$ charge. Apart from the $A_4\times SU(5)$ assignments of 
$F\sim (3,\overbar{5})$, $T_i\sim (1,{10})$, $N^c_i \sim (1,{1})$, 
under $\mathbb{Z}_9$ they transform as $F\sim0$, $T_i\sim (5,7,0)$, $N^c_i\sim (7,8)$.
Unlike the rest of the quarks and leptons, the right-handed neutrinos are further charged under $\mathbb{Z}_{6}$ (as are some of the symmetry breaking scalars). 

In Table \ref{ta:SMF} we also display the six Higgs superfields, generally denoted $H$ (but also $\Lambda$) which serve to break the $SU(5)$ gauge symmetry. 
The two light MSSM Higgs doublet superfields $H_u$ and $H_d$ will emerge from $H_5$ and a mixture of $H_{\bar{5}}$ and $H_{\overbar{45}}$ by a mechanism discussed later. The superfield $\xi$ which breaks $\mathbb{Z}_{9}$ is particularly central to this theory, as it is responsible for both right-handed neutrino masses and the up-type quark mass hierarchy. Finally we have the $\theta_i$ superfields which break $\mathbb{Z}_{6}$ and help to control Dirac neutrino masses, and nine $A_4$ breaking triplet flavons generally denoted $\phi$, with various vacuum alignments, responsible for large lepton mixing.

With these assignments, only the top quark gets a mass from a renormalisable Yukawa coupling $H_5T_3T_3$ 
(which has $\mathbb{Z}_4^R$ charge 2 as required for an allowed superpotential term). 
All the other quark and lepton Yukawa couplings must arise through higher order terms. 
This provides the basic reason why most of the SM (or strictly MSSM) Yukawa couplings appear to be so small.
The observed hierarchy of Yukawa couplings between the three families will be explained via 
a discrete $\mathbb{Z}_{9}$ version of the Froggatt-Nielsen mechanism
\cite{Froggatt:1978nt}, with powers of the low VEV of $\xi$ controlling the hierarchy in the up-type quark sector, and also, in part,
the smallness of the down quark and electron.

In order to enhance predictivity we need the messengers listed in Table \ref{ta:Mess}, which is the price we pay for having a renormalisable theory at the GUT scale. We denote these superfields either as fermion messengers, $X_i$, or scalar messengers, $\Sigma_i$, depending on
whether they carry similar quantum numbers to, respectively, the quarks and leptons
(with odd $\mathbb{Z}_4^R$ charge) or the symmetry breaking scalars 
(with even $\mathbb{Z}_4^R$ charge). The fermion messengers $X_i$ carry similar quantum numbers to down-type quarks and charged leptons (and neutrinos). Scalar messengers $\Sigma_i$ have quantum numbers similar to $H_5$ (the superfield that gives the top quark a renormalisable mass term). The $\Sigma_i$ messengers do not get VEVs, which means we need not consider the effect of diagrams with $\Sigma_i$ superfields in external legs to the masses of SM fermions.

The messengers group themselves in pairs of two superfields with a renormalisable bare mass coupling which respects all the symmetries. Their masses are therefore expected to be at or around the GUT scale.
Although there will be in general distinct masses for different pairs, for simplicity and because they are all expected to be at a similar mass scale, we take masses of all such pairs to be $M$ and set it equal to the GUT scale
in our numerical estimates. We emphasise that the successful predictions of the model
in the lepton sector (namely predicting the PMNS matrix) is independent of the specific values of these 
mass parameters.

\subsection{Up quarks \label{sec:u}}

\begin{figure}[ht]
	\begin{subfigure}[b]{0.17\textwidth}
		\centering
		\includegraphics[scale=0.4]{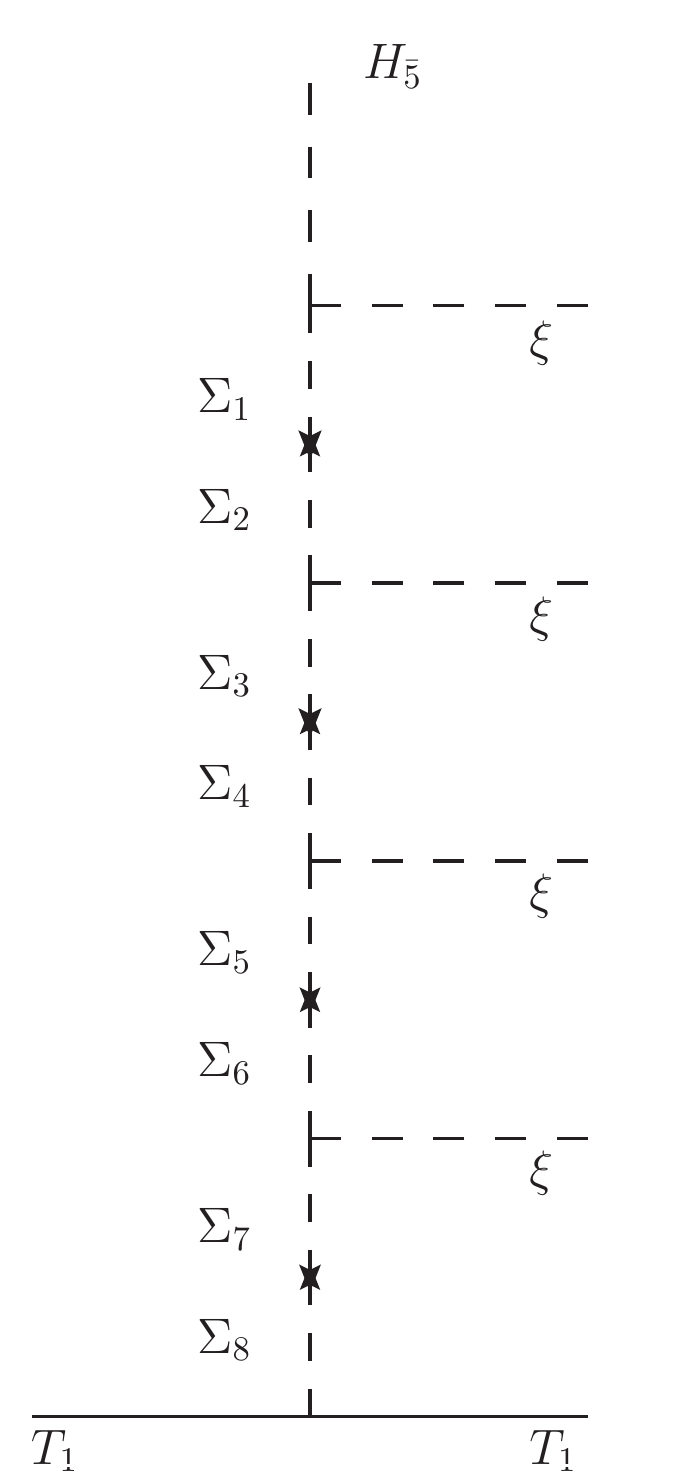}
	\end{subfigure}%
	\begin{subfigure}[b]{0.17\textwidth}
		\centering
		\includegraphics[scale=0.4]{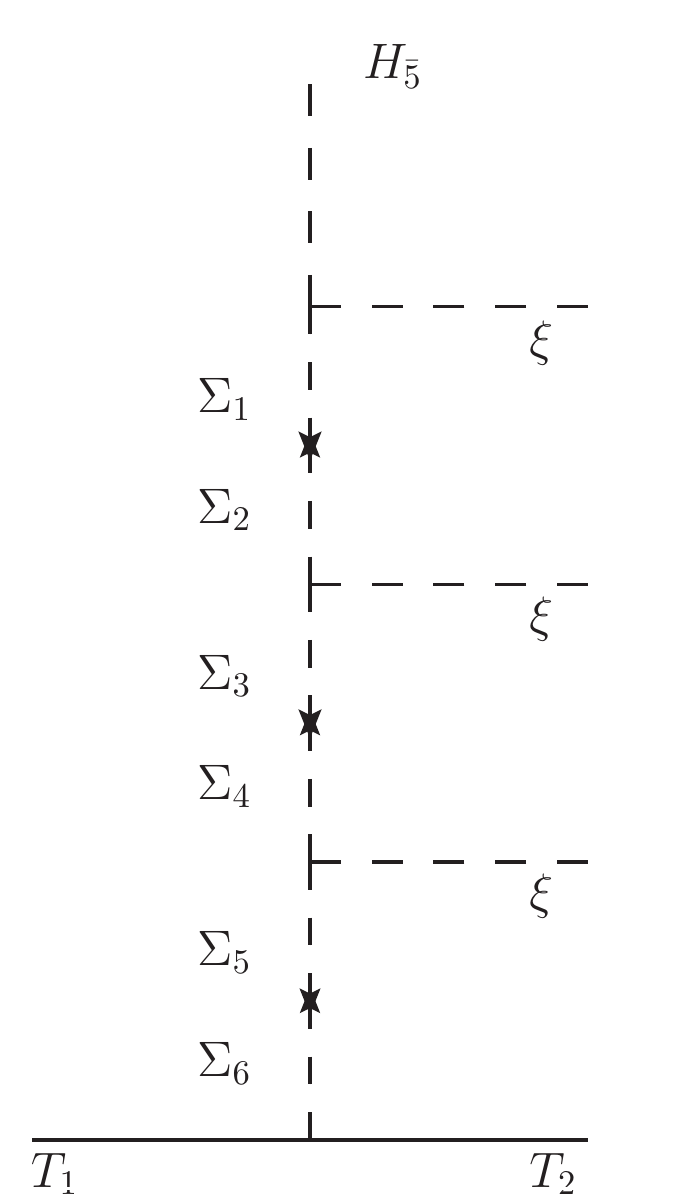}
	\end{subfigure}%
	\begin{subfigure}[b]{0.17\textwidth}
		\centering
		\includegraphics[scale=0.4]{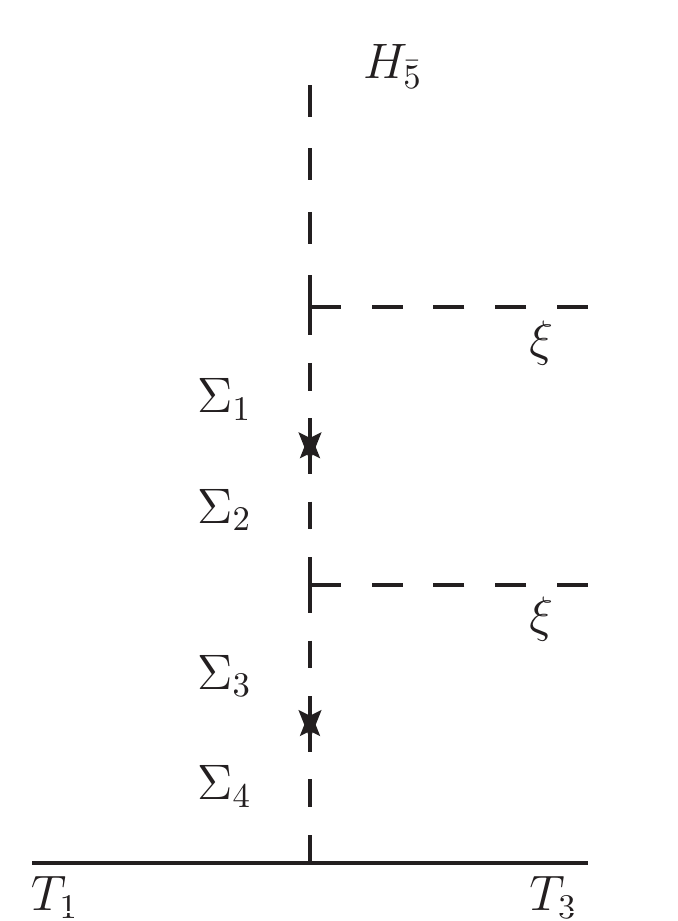}
	\end{subfigure}%
	\begin{subfigure}[b]{0.17\textwidth}
		\centering
		\includegraphics[scale=0.4]{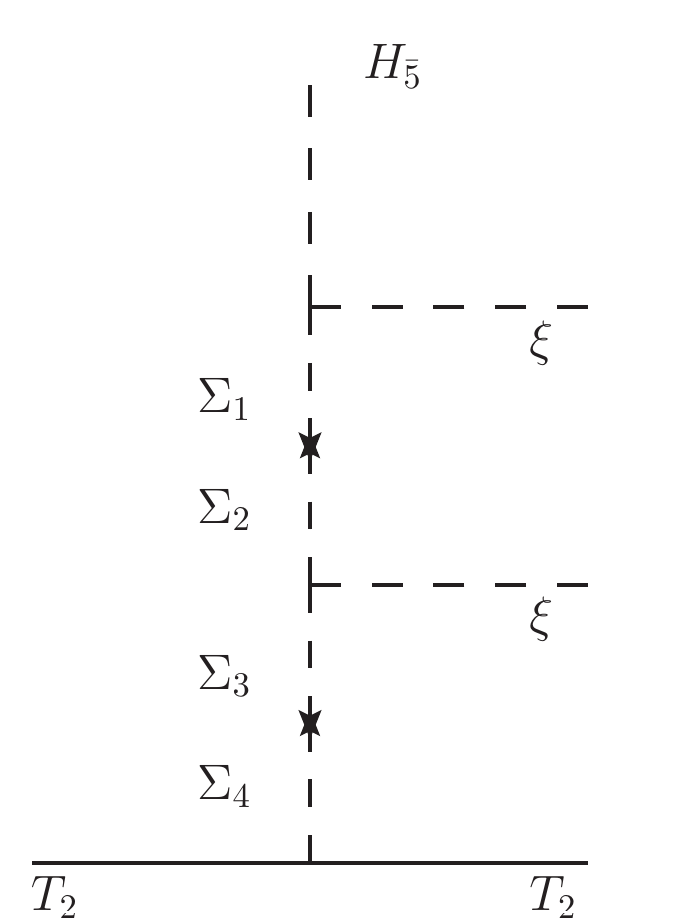}
	\end{subfigure}%
	\begin{subfigure}[b]{0.17\textwidth}
		\centering
		\includegraphics[scale=0.4]{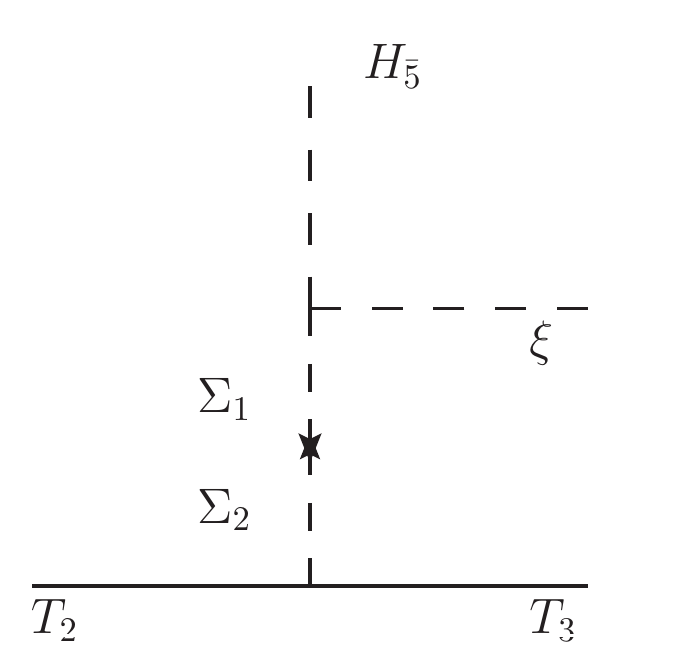}
	\end{subfigure}%
	\begin{subfigure}[b]{0.17\textwidth}
		\centering
		\includegraphics[scale=0.4]{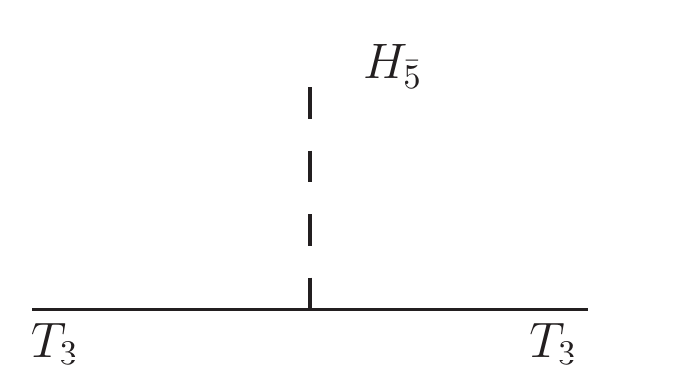}
	\end{subfigure}%
         \caption{Diagrams responsible for the masses and mixings of the up-type quarks.}
	\label{fig:upmass}
\end{figure}

Apart from the top quark mass, which originates from a lowest order Yukawa coupling,
the remaining up-type quark Yukawa couplings appear from higher order terms that result from combining several renormalisable terms involving $\Sigma_i$ messengers and the GUT singlet superfield $\xi$. 
To be precise, the up-type quark Yukawa couplings arise from $\Sigma_i$ messenger tower diagrams shown in Fig.~\ref{fig:upmass}.
For example, the most suppressed coupling arises from the first diagram in Fig.~\ref{fig:upmass}. Other less suppressed couplings arise from the diagrams where at the base one has the respective $T_i T_j$, with a shorter tower leading up to $H_5$. The least suppressed coupling, the renormalisable $H_5 T_3 T_3$ operator responsible for the top quark mass, is the last diagram in Fig.~\ref{fig:upmass}.

The effective superpotential responsible for the up-type Yukawa couplings is
\begin{equation}
	W_{\mathrm{up}} = u_{ij} H_5 T_i T_j \left( \frac{\xi}{M} \right)^{n_{ij}}.
\end{equation}
The resulting symmetric Yukawa matrix for up-type quarks is
\begin{equation}
Y_{ij}^u = u_{ij} \left(\frac{\braket{\xi}}{M}\right)^{n_{ij}} \sim \pmatr{\tilde{\xi}^4 & \tilde{\xi}^3 & \tilde{\xi}^2 \\ &\tilde{\xi}^2 & \tilde{\xi} \\& & 1},
\label{upYuk}
\end{equation}
where $ \tilde{\xi} = \braket{\xi}/M \sim 0.1 $. The explicit form of $ Y^u $ is given in Eq.~\ref{eq:yufull} and includes the coefficients $ u_{ij} $, which are $ \mathcal{O}(1) $ and, by enforcing CP conservation at the GUT scale, necessarily real. Thus, the hierarchy of the up quark masses as well as the CKM mixing angles are given by powers of $\tilde{\xi}$. Due to the structure of this matrix, any phase introduced by $\braket{\xi}$ can be reabsorbed by appropriate redefinition of the three $T_i$ fields, so $Y^u$ does not contain a source of CP violation.

\subsection{Down quarks, charged leptons and flavons \label{sec:dcl}}

When considering the Yukawa structures of down quarks and charged leptons we must inevitably discuss $A_4$ triplet flavons.%
\footnote{As a point of terminology, we refer to as ``flavons'' any superfields that are GUT singlets transforming non-trivially under the family symmetry and that get VEVs. In particular not only $A_4$ but strictly speaking also $\mathbb{Z}_9$ and $\mathbb{Z}_{6}$ are family symmetries, so we also refer to $\xi$ as a ``flavon''.}
The assignments of all the flavons under the family symmetries appear in Table \ref{ta:SMF}. Indeed, since the three families of $F$ transform as a triplet of $A_4$ (see Table \ref{ta:SMF}), all $T_i H_{\bar 5}F$ terms require a contraction with at least one $A_4$ triplet flavon to be invariant.

\begin{figure}[ht]
	\centering
	\begin{subfigure}{0.5\textwidth}
		\centering
		\includegraphics[scale=0.45]{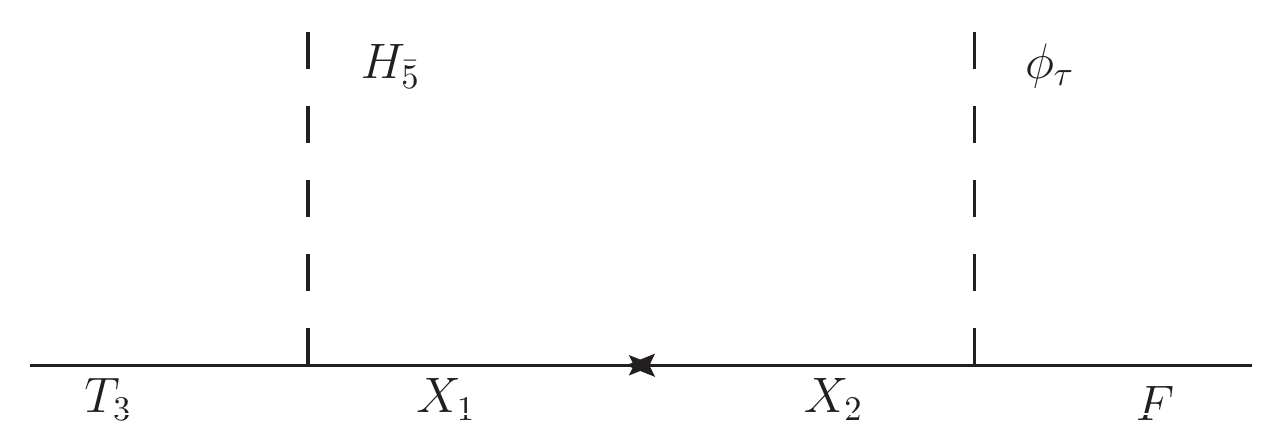}
		\caption{}
	\end{subfigure}%
	\begin{subfigure}{0.5\textwidth}
		\centering
		\includegraphics[scale=0.45]{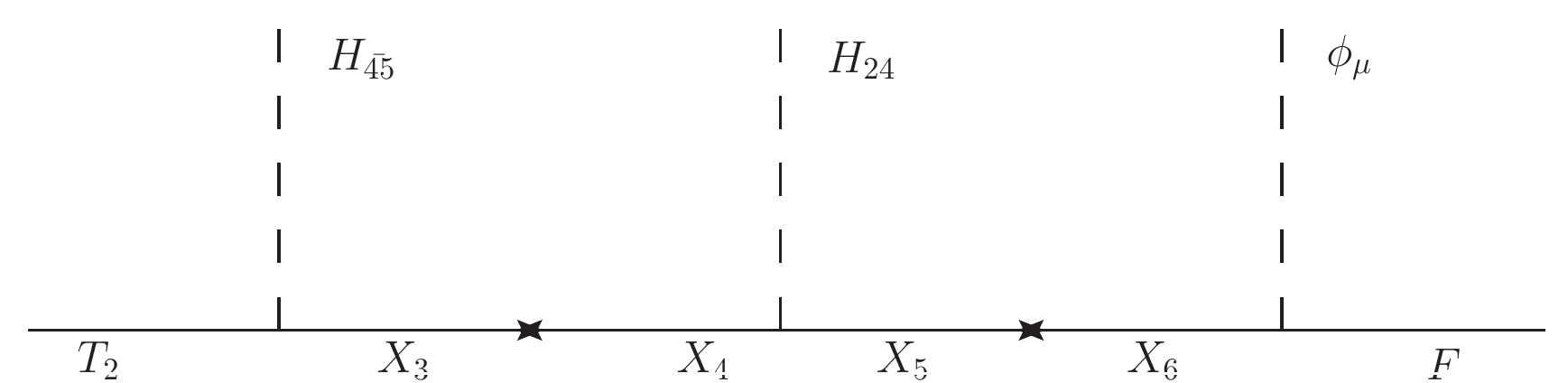}
		\caption{}
	\end{subfigure}
	\begin{subfigure}{0.5\textwidth}
		\centering
		\includegraphics[scale=0.45]{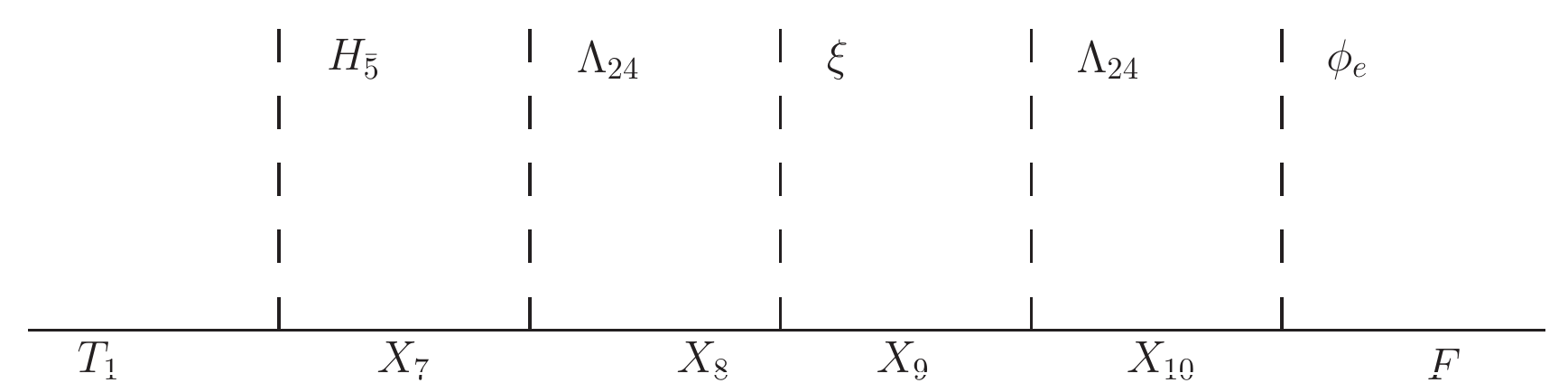}
		\caption{}
	\end{subfigure}%
	\begin{subfigure}{0.5\textwidth}
		\centering
		\includegraphics[scale=0.45]{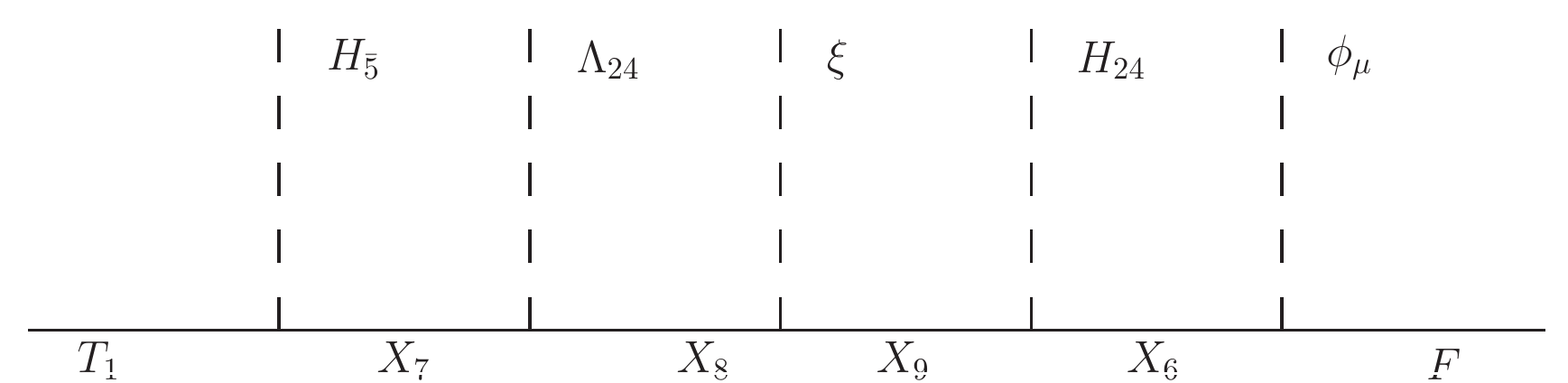}
		\caption{}
		\label{fig:ydoffdiag}
	\end{subfigure}
	\caption{Diagrams responsible for the masses of the down-type quarks and charged leptons.}
	\label{fig:charmass}
\end{figure}

From the diagrams shown in Fig.~\ref{fig:charmass}, integrating out the fermion messengers $X$,
which acquire large masses as a result of either explicit mass terms or GUT scale Higgs VEVs,
we obtain effective operators of the form
\begin{equation}
	W_{\mathrm{down}} = d_{33} T_3 \frac{H_{\bar 5}\phi_\tau}{M}F + d_{22} T_2\frac{H_{\overbar{45}}H_{24}\phi_\mu}{M^2}F + d_{11} T_1\frac{H_{\bar 5}\xi\phi_e}{\braket{\Lambda_{24}}^2}F + d_{12} T_1\frac{H_{\bar 5}\xi\phi_\mu}{\braket{\Lambda_{24}}\braket{H_{24}}}F ,
\label{eq:dcl}
\end{equation}
where $ d_{ij} $ are $ \mathcal{O}(1) $ couplings.
The light MSSM doublet $H_d $ is a combination of the doublets inside $ H_{\bar{5}} $ and $ H_{\overbar{45}} $, as discussed in Section~\ref{sec:splitting}, hence the $d_{22}$ term also leads to a relevant Yukawa coupling.
The alignment of the respective flavon VEVs of $\phi_{e,\mu,\tau}$ (discussed in Section \ref{sec:A4}) is 
\begin{equation}
	\braket{\phi_e} = v_{e} \pmatr{1\\0\\0} \qquad \braket{\phi_\mu} = v_{\mu} \pmatr{0\\1\\0} \qquad \braket{\phi_\tau} = v_{\tau} \pmatr{0\\0\\1},
\end{equation}
such that, apart from $d_{12}$, the contraction appearing with $T_{1,2,3}$ isolates the respective $F_{1,2,3}$ family. This would lead to diagonal Yukawa structures if not for the additional term connecting $T_1 (\phi_\mu F)$ (see Fig.~\ref{fig:ydoffdiag}). 

The resulting effective Yukawa matrices are, schematically:
\begin{equation} 
	Y^d_{LR} \sim Y^e_{RL} \sim \pmatr{ \dfrac{\braket{\xi} v_{e}}{v_{\Lambda_{24}}^2} & \dfrac{\braket{\xi} v_{\mu}}{{v_{\Lambda_{24}}} {v_{H_{24}}}} & 0 \\[2ex] 0&\dfrac{{v_{H_{24}}} v_{\mu}}{M^2} & 0 \\[2ex] 0& 0& \dfrac{v_{\tau}}{M}},
\label{downYuk}
\end{equation}
where $ v_{\Lambda_{24}} $ and $ v_{H_{24}} $ are the respective VEVs of $ \Lambda_{24} $ and $ H_{24} $ (given in Eq.~\ref{eq:su5vevs}), and we include the subscripts $LR$ to emphasise the role of the off-diagonal term to left-handed mixing from $Y^d$. The off-diagonal term in $Y^e$ also provides a tiny contribution to left-handed charged lepton mixing $ \theta_{12}^e \sim m_e/m_\mu $ which may safely be neglected. It also introduces CP violation to the CKM matrix via the phase of $\braket{\xi}$. 

Furthermore, because the underlying renormalisable theory is known, 
the diagrams in Fig.~\ref{fig:charmass} are the only contributions for each family. The $SU(5)$ contractions and associated CG coefficients appearing for each family are unique \cite{Antusch:2009gu,Antusch:2014poa}. 
With the GUT scale symmetry breaking as discussed in Section~\ref{sec:GUT},
each of the scalars here get a VEV with the group structure:
\begin{align}
\begin{split}
	\braket{H_{\bar 5}}^a 			&=\delta^a_5~ v_d/\sqrt{2}\\
	\braket{H_{\overbar{45}}}^{ab}_c 	&=(\delta^{[a}_c-\delta^{[a|}_5\delta^5_c-4\delta^{[a|}_4\delta^4_c)\delta^{b]}_5~v_d/\sqrt{2} \\
	\braket{H_{24}}^a_b 			&=\mathrm{diag}(2,2,2,-3,-3)~v_{H_{24}} \\
	\braket{\Lambda_{24}}^a_b 		&=\mathrm{diag}(2,2,2,-3,-3)~v_{\Lambda_{24}},
\end{split}
\label{eq:su5vevs}
\end{align}
where the indices run $a,b,c=1,...,5$. This leads to the GUT scale prediction:
\begin{equation}
\dfrac{Y^e_{33}}{Y^d_{33}} = 1, \qquad \dfrac{Y^e_{22}}{Y^d_{22}}=\dfrac{9}{2}, \qquad \dfrac{Y^e_{11}}{Y^d_{11}} = \dfrac{Y^e_{21}}{Y^d_{12}} = \dfrac{4}{9}.
\end{equation}
The explicit forms of $ Y^d$ and $ Y^e$, including CG and $ d_{ij} $ coefficients, are given 
later in Eq.~\ref{eq:ydfull} and Eq.~\ref{eq:yefull}, respectively.

\subsection{Neutrinos and CSD3 \label{sec:neutrinos}}

\begin{figure}[ht]
	\centering
	\begin{subfigure}{0.5\textwidth}
		\centering
		\includegraphics[scale=0.6]{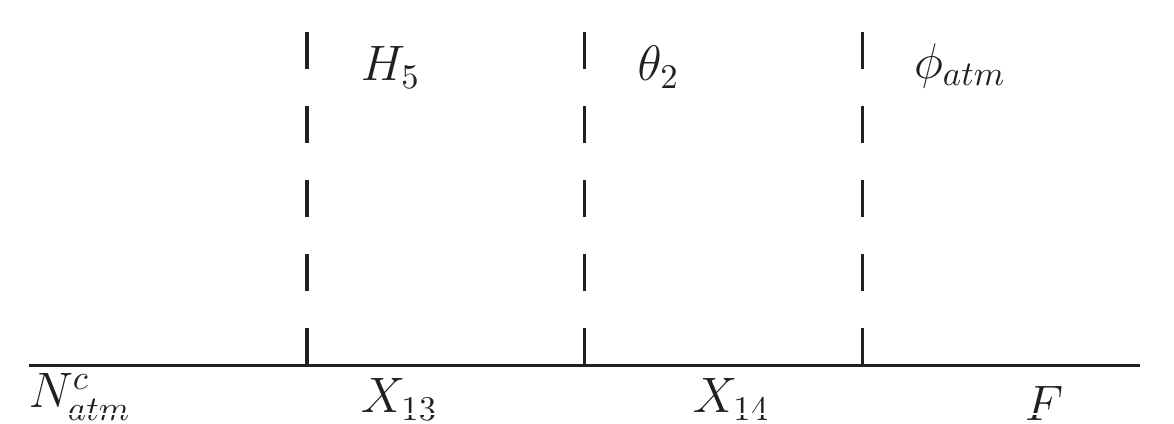}
		\caption{}
	\end{subfigure}%
	\begin{subfigure}{0.5\textwidth}
		\centering
		\includegraphics[scale=0.6]{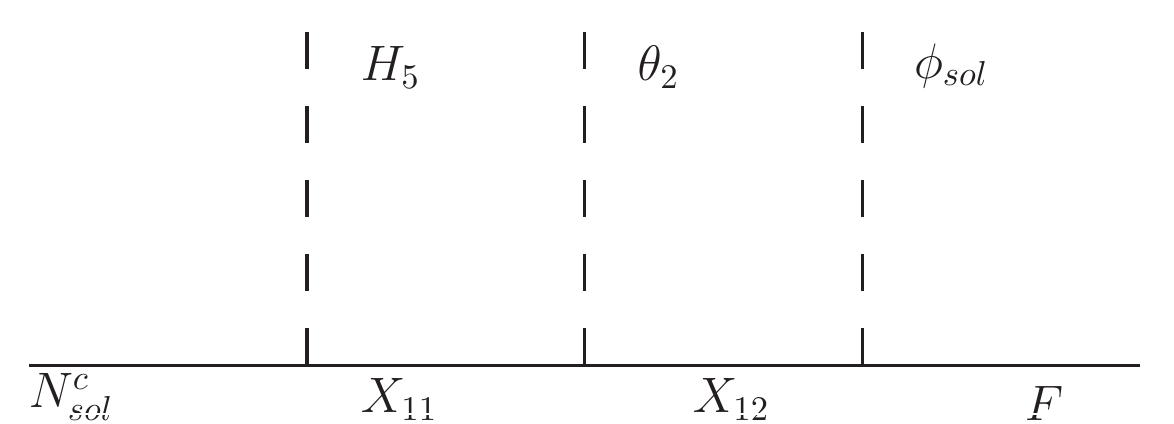}
		\caption{}
	\end{subfigure}
		\begin{subfigure}{0.5\textwidth}
		\centering
		\includegraphics[scale=0.6]{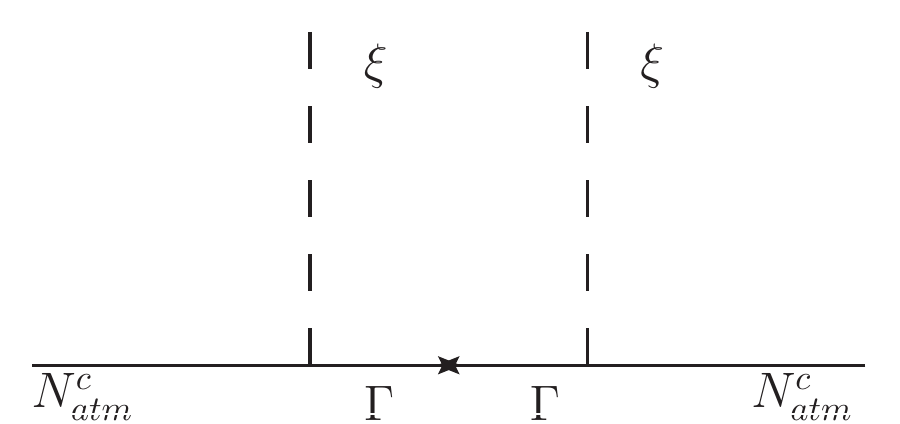}
		\caption{}
	\end{subfigure}%
	\begin{subfigure}{0.5\textwidth}
		\centering
		\includegraphics[scale=0.6]{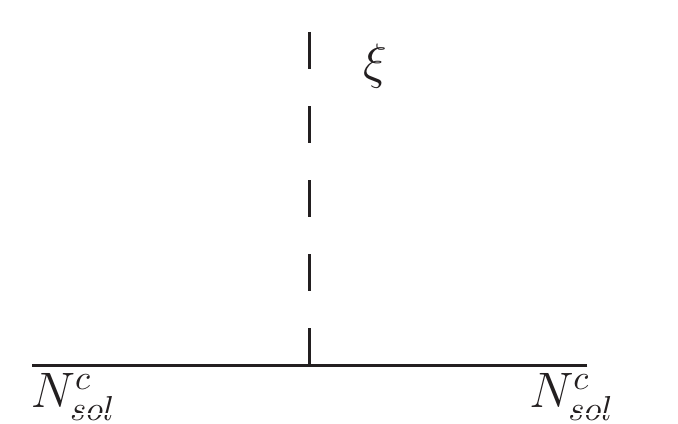}
		\caption{}
	\end{subfigure}
	\caption{Renormalisable diagrams leading to neutrino effective terms. Diagrams (a) and (b) are responsible for neutrino Yukawa terms (leading to Dirac masses) while (c) and (d) give right-handed neutrino Majorana mass terms.}
	\label{fig:Dirac}
\end{figure}

In order to obtain the CSD3 vacuum alignment in this model we couple the neutrinos to a set of flavons distinguished by the $\mathbb{Z}_{6}$ symmetry. Of the superfields in Table~\ref{ta:SMF}, only the right-handed neutrinos and some of the flavons are charged under this symmetry. For clarity, we relabel two of the flavon fields 
as $ \phi_{\mathrm{atm}}  \equiv \phi_3 $ and $  \phi_{\mathrm{sol}} \equiv \phi_4$, to highlight their role in producing neutrino mixing. We also write 
$N_\mathrm{atm}^c \equiv N_1^c$ to denote the right-handed neutrino that dominantly leads to the atmospheric neutrino mass, and $N_\mathrm{sol}^c \equiv N_2^c$ as that which contributes mainly to the solar neutrino mass. 
The relevant terms in the superpotential giving neutrino masses are thus
\begin{equation}
	W_\nu = y_1 H_{5} F\frac{\phi_\mathrm{atm}}{\braket{\theta_2}} N_\mathrm{atm}^c 
	+ y_2 H_{5} F\frac{\phi_\mathrm{sol}}{\braket{\theta_2}} N_\mathrm{sol}^c 
	+ y_3 \frac{\xi^2}{M} N_\mathrm{atm}^c N_\mathrm{atm}^c 
	+ y_4 \xi N_\mathrm{sol}^c N_\mathrm{sol}^c.
\label{eq:neutrinomassW}
\end{equation}
The flavons $\phi_\mathrm{atm}$ and $ \phi_\mathrm{atm} $ gain VEVs according to the CSD3 alignment:
\begin{equation}
	\braket{\phi_{\mathrm{atm}}} = v_{\mathrm{atm}} \pmatr{0\\1\\1} \qquad\qquad \braket{\phi_{\mathrm{sol}}} = v_{\mathrm{sol}} \pmatr{1\\3\\1},
\end{equation}
where $v_{\mathrm{atm}}$ and $v_{\mathrm{sol}}$ are generally complex. Denoting the phases of VEVs as $\rho_i = \arg (v_i)$, only the relative phase $ \relphase $ between the VEVs is physically relevant, and is constrained to a discrete set of values, as discussed in Section \ref{sec:phasefixing}. The flavon $\xi$ (already responsible for the up quark masses) is also acting as a Majoron by generating hierarchical right-handed neutrino masses. At the effective level, the Dirac terms result from coupling the neutrinos (and $H_5$) to $\phi_\mathrm{atm}$ and $\phi_\mathrm{sol}$ via the flavon $\theta_2$ (an $A_4$ singlet carrying $\mathbb{Z}_{6}$ charge). The corresponding diagrams with associated messengers appear in Fig~\ref{fig:Dirac}.

In turn, the Majorana mass term for $N_\mathrm{atm}^c $ is also non-renormalisable and we refer to the superfield $\Gamma$ as the respective messenger. It couples only to $N_\mathrm{atm}^c $ and simply provides the non-renormalisable mass term for $N_\mathrm{atm}^c $, suppressed relative to the mass of $N_\mathrm{sol}^c $. As $\Gamma$ has the quantum numbers of a third right-handed neutrino, one can also consider this field as mediating a double seesaw mechanism, responsible for the $N_\mathrm{atm}^c $ mass. The mixing term $\frac{\xi^6}{M^5} N_{\mathrm{atm}}^c N_{\mathrm{sol}}^c$, though allowed by the symmetries, is absent as there is no combination of messengers able to produce it.

We write $\braket{\xi} = |\xivev | e^{i\xiphase}$, where $ \xiphase $ is chosen from a discrete set of available phases, discussed in Section~\ref{sec:GUTbreaking} (see Eq.~\ref{eq:GUTVEVs}). 
This phase originates from the spontaneous CP violation of a discrete Abelian symmetry \cite{Antusch:2011sx,Ross:2004qn}, in our case the $\mathbb{Z}_9$.
We will now show that $ \xiphase $ and $ \relphase $ fix the relative phases within the effective neutrino mass matrix and consequently the leptonic mixing angles. 

In a Supersymmetric (SUSY) model, the relevant terms in the superpotential giving neutrino masses, in the diagonal charged lepton basis, are
\begin{equation}
	W_\nu = y_{\mathrm{atm}}^i H L_i N_\mathrm{atm}^c 
	+ y_{sol}^i H L_i N_\mathrm{sol}^c 
	+ M_{\rm atm} N_\mathrm{atm}^c N_\mathrm{atm}^c 
	+ M_{\rm sol} N_\mathrm{sol}^c N_\mathrm{sol}^c,
\label{eq:neutrinomassW2}
\end{equation}
where $L_i$ are three families of lepton doublets and the (CP conjugated) right-handed neutrinos $N_{\rm atm}^c$ and $N_{\rm sol}^c$ with real positive masses $M_{\rm atm}$ and $M_{\rm sol}$ do not mix.

The structure of $\lambda^\nu$ is determined by the vacuum alignments of $\phi_\mathrm{atm}$ and $\phi_\mathrm{sol}$. The Dirac and Majorana matrices, derived by comparing the superpotential terms in Eqs.~\ref{eq:neutrinomassW} and \ref{eq:neutrinomassW2} are
\begin{equation}
	\lambda^\nu = \pmatr{0&b\\a&3b\\a&b} \qquad\qquad M^c = \pmatr{\dfrac{y_3\braket{\xi}^2}{M}&0\\[2ex] 0&y_4\braket{\xi}},
	\label{seesaw}
\end{equation}
where $a = y_1 v_{\textrm{atm}}/\braket{\theta_2}$ and $b = y_2 v_{\textrm{sol}}/\braket{\theta_2}$. 

For the see-saw mechanism we shall introduce a different convention for Yukawa and Majorana masses.
The Yukawa matrices $Y^{e}$, $Y^{\nu}$ are defined in a LR convention by
\begin{equation}
	\mathcal{L}^{LR} = -H^dY^e_{ij}\overline{L}^i_L e^j_R - H^u Y^{\nu}_{ij} \overline{L}^i_L  \nu^{i}_R + \mathrm{h.c.},
\end{equation}
where $i,j=1,2,3$ label the three families of lepton doublets $L_i$, right-handed charged leptons $e^j_R$ and right-handed neutrinos $\nu_R^j$ below the GUT scale; $H^u, H^d$ are the Higgs doublets which develop VEVs $v_u,\,v_d$. The physical effective neutrino Majorana mass matrix $m^{\nu}$ is determined by the seesaw mechanism,
\begin{equation}
	m^{\nu} = v_u^2 Y^{\nu} M^{-1}_{R} Y^{\nu \mathrm{T}},
\label{eq:seesaw}
\end{equation}
where the light Majorana neutrino mass matrix $m^\nu$ is defined%
\footnote{The conventions for $ Y^{\nu,e} $ and $m^{\nu}$ differ, respectively, by overall Hermitian and overall complex conjugation compared to those used in the Mixing Parameter Tools package \cite{Antusch:2005gp}, which was used when performing global fits.}
by \( \mathcal{L}^{LL}_\nu = -\tfrac{1}{2} m^\nu \overline{\nu}_{L} \nu^{c}_L + \mathrm{h.c.} \), while the heavy right-handed Majorana neutrino mass matrix $M_R$ is defined by \(\mathcal{L}^{RR}_\nu = -\tfrac{1}{2} M_{R} \overline{\nu^c}_{R} \nu_{R} + \mathrm{h.c.} \).

There is a simple dictionary between the seesaw basis and the SUSY basis, as follows:
compared to the SUSY basis in Eq.~\ref{eq:neutrinomassW} used in leptogenesis calculations we see that $Y^{\nu}=(\lambda^{\nu})^*$, while $M_R=(M^c)^*$. 
Hence the neutrino matrices become, in the seesaw basis,
\begin{equation}
	Y^{\nu} = (\lambda^{\nu})^*=\pmatr{0&b^*\\a^*&nb^*\\a^*&(n-2)b^*} \qquad
	M_R =(M^c)^*= \pmatr{\left(\dfrac{y_3\braket{\xi}^2}{M}\right)^*&0\\[2ex] 0&(y_4\braket{\xi})^*}.
	\label{seesawy}
\end{equation}

Seesaw produces the effective neutrino mass matrix
\begin{equation}
	m^\nu = m_a \pmatr{0&0&0\\0&1&1\\0&1&1} 
			+ m_b e^{i\physphase} \pmatr{1&3&1\\3&9&3\\1&3&1} ,
\label{eq:mnu}
\end{equation}
where $m_a = v_u^2 |a|^2 / (y_3|\xivev|^2 /M)$ and $m_b = v_u^2 |b|^2 / (y_4|\xivev|)$ and we have multiplied throughout by an overall phase which we subsequently drop, keeping only the (physical) relative phase
\begin{equation}
\physphase\equiv -\xiphase+2(\relphase) ,
\label{eta}
\end{equation}
where we recall the above definitions of phases, 
\begin{equation}
	\xiphase \equiv \mathrm{arg}(\braket{\xi}) , \qquad \relphase \equiv \mathrm{arg}(v_{\mathrm{atm}} v_{\mathrm{sol}}^\ast) ,
	\label{varphi}
\end{equation}
and that CP conservation at high energies ensures that $y_i$ and $M$ are real. By arguments given in Section \ref{sec:phasefixing} and Section \ref{sec:GUTbreaking}, we can restrict the physical phase $ \physphase $ to a discrete choice, namely one of the nine complex roots of unity. The values $\physphase= \pm 2\pi/3$ are preferred by CSD3 \cite{King:2013iva,Bjorkeroth:2014vha}. Note that the model predicts a normal neutrino mass hierarchy,  namely $m_3>m_2\gg m_1=0$, which will be tested in the near future. 

The sign of $ \physphase $ has phenomenological significance, as it fixes the leptonic Dirac phase $ \delta^l $. Specifically, a positive $ \physphase $ uniquely leads to negative $ \delta^l $, and \emph{vice versa} \cite{Bjorkeroth:2014vha}. As experimental data hints at $ \delta^l \sim -\pi/2 $, the \emph{a posteriori} preferred solution has positive $ \physphase =+2\pi/3$. The sign of $ \physphase $ also has cosmological significance, as discussed in Section \ref{sec:link}. For example a positive $ \physphase =+2\pi/3$, together with the requirement that baryon asymmetry is positive, implies that the lightest right-handed neutrino should be $N_1^c=N_\mathrm{atm}^c $, while $N_2^c=N_\mathrm{sol}^c $ should be somewhat heavier, which is the natural ordering in our model.

\subsection{Full parameter fit \label{sec:fit}}
The structure of the Yukawa matrices and neutrino mass matrix is set by the theory, up to $\mathcal{O}(1)$ coefficients. The VEVs of the fields $ \xi $, $\Lambda_{24}$ and $H_{24}$ are at or near the GUT scale, but otherwise undetermined. This freedom coincides with the choice of coefficients in the Yukawa matrices, providing no extra degrees of freedom in the determination of the Yukawas other than to provide the appropriate scale. The same is true for the flavon fields $\phi_e$, $\phi_\mu$ and $\phi_\tau$, which provide the necessary hierarchy in the down-quark and charged lepton Yukawa sector. 

The neutrino matrix $ m^\nu $ is given in Eq.~\ref{eq:mnu}. Letting $ v_f $ represent the VEV of a field $ f $, the Yukawa matrices are as follows:
\begingroup
\begin{align}
	Y^u &= \pmatr{	u_{11} |\tilde{\xi}^4| & u_{12} |\tilde{\xi}^3| & u_{13} |\tilde{\xi}^2| \\[2ex]
				u_{12} |\tilde{\xi}^3| & u_{22} |\tilde{\xi}^2| & u_{23} |\tilde{\xi}| \\[2ex]
				u_{13} |\tilde{\xi}^2| & u_{23} |\tilde{\xi}| & u_{33}
		} \label{eq:yufull} \\ \nonumber \\
	Y^d &=\frac{1}{\sqrt{2}} \pmatr{\dfrac{1}{4} d_{11} \dfrac{|\xivev v_e|}{|v_{\Lambda_{24}}|^2}  
	& d_{12}  \dfrac{ | \xivev v_\mu |}{|v_{\Lambda_{24}} v_{H_{24}}|}  e^{i \zeta}   & 0 \\[2.5ex]
				0 & 2 d_{22}  \dfrac{| v_{H_{24}} v_\mu | }{M^2} & 0 \\[2.5ex]
				0 & 0 & d_{33} \dfrac{|v_\tau |}{M} 
		} \label{eq:ydfull}\\ \nonumber \\
	Y^e &=\frac{1}{\sqrt{2}}\pmatr{\dfrac{1}{9} d_{11} \dfrac{|\xivev v_e|}{|v_{\Lambda_{24}}|^2} & 0 & 0 \\[2.5ex]
			d_{12} \dfrac{|\xivev v_\mu|}{|v_{\Lambda_{24}} v_{H_{24}}|} e^{i \zeta} & 9 d_{22} \dfrac{|v_{H_{24}} v_\mu|}{M^2} & 0 \\[2.5ex]
			0 & 0 & d_{33} \dfrac{|v_\tau|}{M} .
	} \label{eq:yefull}
\end{align}
\endgroup
As already remarked, the phases in $Y^u$ from powers of $\braket{\xi} = |\xivev |e^{i\xiphase}$ 
can be removed by field redefinition. 
Without loss of generality we have rephased fields such that the only phase appearing in
$ Y^d $ and $ Y^e$ is the phase $ \zeta $ as shown in Eqs.~\ref{eq:ydfull}, \ref{eq:yefull}, so all quark CP violation
originates from the single phase $ \zeta $ appearing in $Y^d_{12}$.
In turn, $ \zeta $ is determined by a combination of phases coming from various field VEVs; more precisely, $\zeta=\rho_\xi-2\rho_{H_{24}}-\rho_{\Lambda_{24}}$.
As long as it is reasonably far from zero, it can produce the necessary CP violation. Different choices of 
$ \zeta $ do not affect the goodness-of-fit, corresponding simply to different but equally valid choices of $ \mathcal{O}(1) $ coefficients. For our fit we choose $ \zeta = \pi/3$.
Note that the corresponding phase in $Y^e_{21}$ does not contribute to leptonic CP violation, since this term does not affect left-handed mixing, to an accuracy of $\mathcal{O}(m_e/m_{\mu})$.

To fit the real coefficients $u_{ij}$, $d_{ij}$, $m_a$ and $m_b$, we propose a function $\chi^2$ that relates the $N$ physical predictions $P_i(\{x\})$ for a given set of input parameters $\{x\}$ to their current best-fit values $\mu_i$ and their associated $1\sigma$ errors, denoted $ \sigma_i $, by
\begin{equation}
	\chi^2 = \sum_{i=1}^N \left(\frac{P_i(\{x\})-\mu_i}{\sigma_i} \right)^2.
	\label{eq:chisq}
\end{equation}
The errors $ \sigma_i $ are equivalent to the standard deviation of the experimental fits to a Gaussian distribution. For most parameters, this is essentially the case, with the exception of the (lepton) atmospheric angle $ \theta^l_{23} $. For a normal hierarchy, the distribution is roughly centered on $ \theta^l_{23} = 45^\circ $, while the best fit value is given by $ \theta^l_{23} = 42.3^\circ $. So as to not overstate the error for $ \theta^l_{23} $, we approximate its distribution by a Gaussian about $ 42.3^\circ $, setting $ \sigma_{\theta^l_{23}} = 1.6^\circ $.

We now wish to minimise $ \chi^2 $. In this analysis, $N = 18$, corresponding to six mixing angles $\theta^l_{ij}$ (neutrinos) and $\theta^q_{ij}$ (quarks), the CKM phase $\delta^q$, nine Yukawa eigenvalues for the quarks and charged leptons, and two neutrino mass-squared differences $\Delta m^2_{21}$ and $\Delta m^2_{31}$. 
In the lepton sector, we use the PDG parametrisation of the PMNS matrix \cite{Beringer:1900zz}
\(U_{\mathrm{PMNS}} = R^l_{23} U^l_{13} R^l_{12} P_\textrm{PDG}  \) in terms of \(s_{ij}=\sin \theta^l_{ij}\), \(c_{ij}=\cos\theta^l_{ij}\), the Dirac CP violating phase \(\delta^l\) and
further Majorana phases contained in  
\(P_\textrm{PDG} = \textrm{diag}(1,e^{i\frac{\alpha_{21}}{2}},e^{i\frac{\alpha_{31}}{2}})\). 
Experimentally, the leptonic phase $ \delta^l $ is poorly constrained at $ 1\sigma $ (and completely unconstrained at $ 3\sigma $), so is not fit, and left as a pure prediction of the model, as are the (completely unconstrained) Majorana phases $ \alpha_{21} $ and $ \alpha_{31} $. As this model predicts only two massive left-handed neutrinos, i.e. $ m_1 = 0 $, one Majorana phase is zero, which we take to be 
$ \alpha_{31} =0$.

\begingroup
\newcommand{\exto}[1]{{\smaller$ \times 10^{#1} $}}
\newcommand{\degr}{$^\circ$}
\begin{table}[!ht]
\centering
\footnotesize
\begin{tabular}{| c | c c | c c |}
\hline
\multirow{2}{*}{\makecell{Parameter \\ {\scriptsize (from \cite{Antusch:2013jca})}}}	& \multicolumn{2}{c|}{\rule{0pt}{3ex}$ \tan \beta = 5 $} & \multicolumn{2}{c|}{$ \tan \beta = 10 $} \\[0.5ex]
\cline{2-5}
				& $ \mu_i $ & $ \sigma_i $ & $ \mu_i $ & $ \sigma_i $ \\
\hline \hline
\rule{0pt}{3ex}%
$\theta^q_{12}$ 		& 13.027\degr 	& 0.0814\degr 		& 13.027\degr 	& 0.0814\degr 	\\
$\theta^q_{13}$ 		& 0.1802\degr 	& 0.0281\degr 		& 0.1802\degr 	& 0.0281\degr 	\\
$\theta^q_{23}$ 		& 2.054\degr 	& 0.384\degr		& 2.054\degr 	& 0.384\degr	\\
$\delta^q$ 			& 69.21\degr	& 6.19\degr 		& 69.21\degr	& 6.19\degr 	\\
\rule{0pt}{3ex}%
$y_u$				& 2.92 \exto{-6} & 1.81 \exto{-6}	& 2.88 \exto{-6} & 1.79 \exto{-6} \\
$y_c$				& 1.43 \exto{-3} & 1.00 \exto{-4}	& 1.41 \exto{-3} & 9.87 \exto{-5} \\
$y_t$				& 5.34 \exto{-1} & 3.41 \exto{-2}	& 5.20 \exto{-1} & 3.15 \exto{-2} \\
\rule{0pt}{3ex}%
$y_d$				& 4.81 \exto{-6} & 1.06 \exto{-6}	& 4.84 \exto{-6} & 1.07 \exto{-6} \\
$y_s$				& 9.52 \exto{-5} & 1.03 \exto{-5}	& 9.59 \exto{-5} & 1.04 \exto{-5} \\
$y_b$				& 6.95 \exto{-3} & 1.75 \exto{-4}	& 7.01 \exto{-3} & 1.78 \exto{-4} \\
\rule{0pt}{3ex}%
$y_e$				& 1.97 \exto{-6} & 2.36 \exto{-8}	& 1.98 \exto{-6} & 2.38 \exto{-8} \\
$y_\mu$			& 4.16 \exto{-4} & 4.97 \exto{-6}	& 4.19 \exto{-4} & 5.02 \exto{-6} \\
$y_\tau$			& 7.07 \exto{-3} & 7.27 \exto{-5}	& 7.15 \exto{-3} & 7.42 \exto{-5} \\
\hline
\end{tabular}
\caption{Best fit values for quark and charged lepton parameters when run to the GUT scale as calculated in \cite{Antusch:2013jca}, with the SUSY breaking scale set at 1 TeV. We have included an overall contribution from threshold corrections corresponding to $ \bar{\eta}_b = -0.24375 $ which affects primarily the $ b $ quark Yukawa coupling $ y_b $. $ \mu_i $ represents the best-fit value and $ \sigma_i $ the error, as defined in Eq.~\ref{eq:chisq}.}
\label{tab:data}
\end{table}
\endgroup

The running of best-fit and error values to the GUT scale are generally dependent on SUSY parameters, notably $\tan \beta$, as well as contributions from SUSY threshold corrections. We extract the GUT scale CKM parameters and all Yukawa couplings (with associated errors) from \cite{Antusch:2013jca} for judicious choices of $ \tan \beta $. In further reference to \cite{Antusch:2013jca}, we choose for the parameter $ \bar{\eta}_b $ parametrising the threshold corrections a value $ \bar{\eta}_b = -0.24375 $; a non-zero value is required primarily to produce a necessary (small) difference in $ b $ and $ \tau $ Yukawa couplings. Experimental neutrino parameters are extracted from \cite{Gonzalez-Garcia:2014bfa}. All data is reproduced in Tables \ref{tab:data} and \ref{tab:lepdata}.

\begingroup
\newcommand{\exto}[1]{{\smaller$ \times 10^{#1} $}}
\newcommand{\degr}{$^\circ$}
\begin{table}[ht]
\centering
\footnotesize
\renewcommand{\arraystretch}{1.5}
\begin{tabular}{| c | r r | }
\hline
\rule{0pt}{4ex}\makecell{Parameter \\ {\scriptsize (from \cite{Gonzalez-Garcia:2014bfa})}}
						& \(\mu_i \pm1\sigma\) & 3\(\sigma\) range \\[1.5ex] \hline\hline
\( \theta^l_{12} \,(^\circ)\)			& 33.48 \(^{+0.78}_{-0.75}\)		& 31.29 \(\rightarrow\) 35.91	\\ 
\( \theta^l_{23} \,(^\circ)\)			& 42.3 \(^{+3.0}_{-1.6}\)		& 38.2 \(\rightarrow\) 53.3 	\\ 
\( \theta^l_{13} \,(^\circ)\)			& 8.5 \(^{+0.20}_{-0.21}\)		& 7.85 \(\rightarrow\) 9.10 	\\ 
\( \delta^l \,(^\circ)\)			& 306 \(^{+39}_{-70}\)			& 0 \(\rightarrow\) 360		\\ [5pt]
\(\dfrac{\Delta m^2_{21}}{10^{-5}}\) eV\(^2\)	& 7.50 \(^{+0.19}_{-0.17}\)		& 7.02 \(\rightarrow\) 8.09	\\ [8pt]
\(\dfrac{\Delta m^2_{31}}{10^{-3}}\) eV\(^2\) 	& +2.457 \(^{+0.047}_{-0.047}\)	& +2.317 \(\rightarrow\) +2.607\\ [5pt]
\hline
\end{tabular}
\caption{Table of current best fits to experimental data for lepton mixing angles and neutrino masses case of normal mass squared ordering taken from \cite{Gonzalez-Garcia:2014bfa}, with 1\(\sigma\) and 3\(\sigma\) uncertainty ranges.}
\label{tab:lepdata}
\end{table}
\endgroup

Minimisation by differential evolution was performed in Mathematica, producing the set of $ \mathcal{O}(1) $ input coefficients and the corresponding physical parameters seen in Table \ref{tab:fit}, with an associated $ \chi^2 = 7.98 $ (for $ \tan \beta = 5 $) and $ \chi^2 = 7.84 $ (for $ \tan \beta = 10 $).

In this fit, the VEVs of $ \xi $, $ \Lambda_{24} $, $ H_{24} $ and the three $ \phi_{e,\mu,\tau} $ are fixed by hand in terms of the scale $ M $, which is taken to be the GUT scale, i.e. $ M \approx 3 \times 10^{16} $ GeV. Similarly, the Higgs doublet VEV enters only implicitly through $ m_a $ and $ m_b $, but is understood to take (at the GUT scale) the value $ v_H = 174 $ GeV. We set
\begin{equation}
\begin{split}	\xivev &= 6 \times 10^{-2} M \\ v_{\Lambda_{24}} &= M \\ v_{H_{24}} &= 3 \times 10^{-1}M \end{split} \qquad \qquad
\begin{split}	v_e &= 10^{-3} M \\ v_\mu &= 10^{-3} M \\ v_\tau &= 5 \times 10^{-2} M \end{split}.
\label{eq:vevfit}
\end{equation}
The value of $ \xivev $ is chosen to accommodate not only the fit to $ Y^u $ parameters but also to control the $ \mu $-term, as discussed in Section \ref{sec:splitting}. Meanwhile the factor $ \sim$3 split between $v_{\Lambda_{24}}$ and $ v_{H_{24}} $ assists in establishing a hierarchy between the $ e $ and $ \mu $ families.
With the above numerical values for the VEVs, we get the following Yukawa matrices in terms only of $ \mathcal{O}(1) $ coefficients and the complex phase $ \zeta $:
\begingroup
\newcommand{\pmatrs}[1]{\begin{pmatrix*}[r] #1 \end{pmatrix*}}
\newcommand{\extom}[1]{{{\scriptstyle\times 10^{#1}}}}
\begin{align}
	Y^u &= \phantom{\frac{1}{\sqrt{2}}} \pmatrs{1.296 \extom{-5} \cdot u_{11} & 2.16 \extom{-4} \cdot u_{12} & 3.6\extom{-3} \cdot u_{13} \\ 2.16 \extom{-4} \cdot u_{21} & 3.6\extom{-3} \cdot u_{22} & 6 \extom{-2} \cdot u_{23} \\ 3.6\extom{-3} \cdot u_{31} & 6\extom{-2} \cdot u_{32} & u_{33}} \\[2ex]
	Y^d &= \frac{1}{\sqrt{2}} \pmatr{ 1.5\extom{-5} \cdot d_{11} \phantom{e^{i \zeta}} & 2 \extom{-4} \cdot d_{12} e^{i \zeta} & 0 \\ 0 & 6 \extom{-4} \cdot d_{22} \phantom{e^{i \zeta}} & 0 \\ 0 & 0 & 5 \extom{-2} \cdot d_{33}} \\[2ex]
	Y^e &= \frac{1}{\sqrt{2}} \pmatr{6.67 \extom{-6} \cdot d_{11} \phantom{e^{i \zeta}} & 0 & 0 \\ \phantom{6.6}2 \extom{-4} \cdot d_{12} e^{i \zeta} & 2.7 \extom{-3} \cdot d_{22}  & 0 \\ 0 & 0 & 5 \extom{-2} \cdot d_{33}}.
\end{align}
\endgroup

\begingroup
\newcommand{\exto}[1]{{\smaller$ \times 10^{#1} $}}
\makeatletter
\newcommand{\thickhline}{%
    \noalign {\ifnum 0=`}\fi \hrule height 1pt
    \futurelet \reserved@a \@xhline
}
\makeatother
\newcommand{\degr}{$^\circ$}

\begin{table}[ht]
\centering
\footnotesize
\begin{tabular}{| c | c | cc | cc |}
\hline
\rule{0pt}{3ex}$ \tan \beta $	& Input & \multicolumn{4}{|c|}{Output} \\[1ex]\hline\hline
\rule{0pt}{3ex}%
\multirow{9}{*}{5}	& 
 \multirow{7}{*}{$\begin{array}{rl}u_{ij}: & \pmatr{0.9566 & 0.7346 & 0.7198 \\ \cdot & 0.5961 & 0.3224 \\ \cdot & \cdot & 0.5435 } \\[6ex] d_{ij}: & \pmatr{2.133 & 0.8363 &  \\  & 1.108 &  \\  &  & 1.021 }\end{array}$} 
 	& $\theta^q_{12}$ 		& 13.027\degr	& $ \theta^l_{12} $ 	& 34.3\degr \\
&	& $\theta^q_{13}$ 		& 0.1802\degr  & $ \theta^l_{13} $ 	& 8.67\degr \\
&	& $\theta^q_{23}$ 		& 2.054\degr	& $ \theta^l_{23} $ 	& 45.8\degr \\
&	& $\delta^q $ 			& 69.18\degr	& $ \delta^l $		& -86.7\degr \\
\rule{0pt}{3ex}%
&	& $y_u$ 			& 2.92 \exto{-6}	& $\Delta m^2_{21}$ 	& 7.38 \exto{-5} eV$^2$ \\
&	& $y_c$ 			& 1.43 \exto{-3}	& $\Delta m^2_{31}$ 	& 2.48 \exto{-3} eV$^2$\\
&	& $y_t$ 			& 5.34 \exto{-1}	& &\\
&	& $y_d$ 			& 4.30 \exto{-6}	& $y_e$ 		& 1.97 \exto{-6} \\
&$ m_a: 26.57 $ meV	& $y_s$ 	& 9.51 \exto{-5}	& $y_\mu$ 		& 4.16 \exto{-4} \\
&$ m_b: 2.684 $ meV	& $y_b$ 	& 7.05 \exto{-3}	& $y_\tau$ 		& 7.05 \exto{-3} \\[1ex]
\thickhline
\rule{0pt}{3ex}%
\multirow{9}{*}{10}	& 
 \multirow{7}{*}{$\begin{array}{rl} u_{ij}:& \pmatr{0.9182 & 0.7087 & 0.6910 \\ \cdot & 0.5768 & 0.3095 \\ \cdot & \cdot & 0.5218} \\[6ex] d_{ij}: & \pmatr{4.236 & 1.661 &  \\  & 2.200 &  \\  &  & 2.034}\end{array}$}  
 	& $\theta^q_{12}$ 		& 13.027\degr	& $ \theta^l_{12} $ 	& 34.3\degr \\
&	& $\theta^q_{13}$ 		& 0.1802\degr  & $ \theta^l_{13} $ 	& 8.67\degr \\
&	& $\theta^q_{23}$ 		& 2.054\degr	& $ \theta^l_{23} $ 	& 45.8\degr \\
&	& $\delta^q $ 			& 69.18\degr	& $ \delta^l $		& -86.7\degr \\
\rule{0pt}{3ex}%
&	& $y_u$ 			& 2.88 \exto{-6}	& $\Delta m^2_{21}$ 	& 7.38 \exto{-5} eV$^2$ \\
&	& $y_c$ 			& 1.41 \exto{-3}	& $\Delta m^2_{31}$ 	& 2.48 \exto{-3} eV$^2$ \\
&	& $y_t$ 			& 5.20 \exto{-1}	& &\\
&	& $y_d$ 			& 4.33 \exto{-6}	& $y_e$ 		& 1.98 \exto{-6} \\
&$ m_a: 26.57 $ meV	& $y_s$ 	& 9.58 \exto{-5}	& $y_\mu$ 		& 4.19 \exto{-4} \\
&$ m_b: 2.684 $ meV	& $y_b$ 	& 7.13 \exto{-3}	& $y_\tau$ 		& 7.13 \exto{-3} \\[1ex]
\hline
\end{tabular}
\caption{Fitted input quark Yukawa coefficients $ u_{ij} $ and $ d_{ij} $ (arranged by their position in the $ Y^u $ and $ Y^d $ matrices, respectively) and neutrino mass parameters $ m_a $ and $ m_b $, and associated physical parameters produced by minimising the function $ \chi^2 $. We choose $ \zeta = \pi/3$.
With $ \tan \beta = 5 $, the fit gives $ \chi^2 = 7.98 $, while $ \tan \beta = 10 $ gives $ \chi^2 = 7.84 $, both very good fits. The largest single contribution to $ \chi^2 $ is from the fit to the atmospheric angle $ \theta_{23}^l $. 
These results are given with $ \eta = +2\pi/3 $. The non-zero Majorana phase is also predicted to be $ \alpha_{21} = 72^\circ$, and is insensitive to $ \tan \beta $, as indeed are all the mixing angles and phases. }
\label{tab:fit}
\end{table}

\endgroup

It is worth reiterating that the neutrino mass matrix phase $ \physphase $ can be forced to admit only phases coming from the nine complex roots of unity, essentially due to spontaneous CP violation with the $\mathbb{Z}_9$ symmetry, where we select $ \physphase = \pm 2\pi/3 $, both of which yield equally good $ \chi^2 $ fits, with only minor adjustments to $ \mathcal{O}(1) $ coefficients. The primary effect is in the prediction of $ \delta^l $; as previously stated, positive $ \physphase $ corresponds to negative $ \delta^l $. As this is preferred by experiment, the results presented in Table \ref{tab:fit} are for $ \physphase =+2\pi/3$. The fit also predicts the Majorana phases $ \alpha_{21} = 72^\circ $ and $ \alpha_{31} = 0 $.

In order to understand the significance of the $ \chi^2 $ fit, and assess the strength of the model overall, it is prudent to enumerate the parameters and predictions of the model. The nominal parameter count at the GUT scale is very large, owing to the diverse field content. However, at the scale where we are able to make predictions, many of these parameters combine to give a constrained set of free parameters that need to be determined. Notably, the VEVs of Higgs and flavon fields such as those given in Eq.~\ref{eq:vevfit} do not constitute true degrees of freedom, as they can be absorbed by redefining other parameters. 

Relevant parameters that require consideration include: six $ u_{ij} $, four $ d_{ij} $, masses $ m_a $ and $ m_b $, phases $ \eta $ and $ \zeta $, the threshold factor $ \bar{\eta}_b $, and $ \tan \beta $, for a total of $N_I=16$ input parameters.
However three of these parameters, namely
$\tan \beta $, $\eta$ and $ \zeta $, are fixed prior to the fit, with the latter two phases restricted to discrete choices, as discussed previously.
Finally, the factor $ \bar{\eta}_b $ affects only the coupling $ y_b $ and is fitted by hand. As mentioned earlier, the model fits $ N = 18 $ observables, including nine Yukawa eigenvalues, two neutrino mass squared differences, six mixing angles and the quark CP phase. In addition the model  predicts 
the leptonic CP phase $ \delta^l $, two Majorana phases (one of which is zero) and a massless physical neutrino.

\section{\secheadmath{A_4} symmetry breaking and the flavon vacuum\label{sec:A4}}

In order to address $A_4$ symmetry breaking we need to address three aspects of the flavon vacuum: what drives some flavons to have VEVs a few orders of magnitude below the GUT scale, what determines their vacuum alignment, and what fixes the relative vacuum phase $\relphase \equiv \mathrm{arg}(v_{\mathrm{atm}} v_{\mathrm{sol}}^{\ast})$ (and consequently the physical phase $ \physphase $). In this section we consider each of these issues in turn.

\subsection{Driving the flavon vacuum expectation values \label{sec:radiative}}
The flavon $\phi$ VEVs are driven by radiative breaking \cite{Ibanez:1982fr} (see e.g. \cite{Ibanez:2007pf} for a recent review). The soft squared mass terms appear as $m_i^2 \phi_i \phi_i^\dagger \mathrm{ln}(\phi_i \phi_i^\dagger/\Lambda_i^2)$ and become negative at the required scales $\Lambda_i < M$, lifting the respective flat direction to a few orders of magnitude below the GUT scale for each flavon: $\langle \phi_i \phi_i^\dagger \rangle \sim \Lambda_i^2$. Hierarchies of VEVs are thus naturally expected due to the logarithmic nature of this mechanism. The precise symmetry breaking scale $\Lambda_i$ for each field $ \phi_i$ depends on otherwise undetermined parameters in the model which are different for each flavon, such as superpotential terms involving messengers.%
\footnote{Examples of such a Yukawa coupling for the neutrino flavons are $\phi_{\mathrm{atm}} F X_{14}$ and $\phi_{\mathrm{sol}} F X_{12}$.}
Therefore a hierarchy for such flavon VEVs is generated and remains stable due to radiative breaking \cite{Greene:1986jb}. On the other hand those for $v_e, v_{\mu}, v_{\tau}$ in Eq.~\ref{eq:vevfit} arise from $F$-terms as discussed later.

\subsection{Flavon vacuum alignment}

\begingroup

\newcommand{\five}{$\textrm{5}$}
\newcommand{\fivebar}{$\bar{\textrm{5}}$}
\newcommand{\fortyfivebar}{$\overbar{45}$}
\newcommand{\onep}{1$^{\prime}$}
\newcommand{\onepp}{1$^{\prime\prime}$}
\newcommand{\fiftybar}{$\overbar{50}$}

\begin{table}
\centering
\footnotesize
\begin{minipage}[b]{0.45\textwidth}
\captionsetup{width=\textwidth}
\centering
\begin{tabular}{| c | c c | c | c | c |}
\hline
\multirow{2}{*}{\rule{0pt}{4ex}Field}	& \multicolumn{5}{c |}{Representation} \\
\cline{2-6}
\rule{0pt}{3ex}		& $A_4$ 	& SU(5) 	& $\mathbb{Z}_9$ & $\mathbb{Z}_6$ & $\mathbb{Z}_4^R$ \\ [0.75ex]
\hline \hline
\rule{0pt}{3ex}%
$Z_1$			& \onepp		& 24		& 0 & 0 & 2 \\
$Z_2$ 			& \onepp		& 24		& 3 & 0 & 2 \\
$Z_3$ 			& \onep		& 24		& 3 & 0 & 2 \\
\rule{0pt}{3ex}%
$\Upsilon_{1}$		& \onep	& 24 	& 7 & 0 & 0 \\
$\Upsilon_{2}$		& \onepp	& 24		& 2 & 0 & 2 \\
$\Upsilon_{3}$		& \onep	& 24		& 5 & 0 & 0 \\
$\Upsilon_{4}$	 	& \onepp	& 24		& 4 & 0 & 2 \\
$\Upsilon_{5}$		& \onep 	& 24	 	& 4 & 0 & 0 \\
$\Upsilon_{6}$		& \onepp 	& 24		& 5 & 0 & 2 \\
$\Upsilon_{7}$		& \onep	& 24		& 2 & 0 & 0 \\
$\Upsilon_{8}$		& \onepp 	& 24		& 7 & 0 & 2 \\ 
\rule{0pt}{3ex}%
$\Upsilon_{9}$		& 1  	 	& 75	 	& 0 & 0 & 0 \\
$\Upsilon_{10}$	& 1	 	& 75		& 0 & 0 & 2 \\
$\Upsilon_{11}$	& 1		& 75		& 6 & 0 & 0 \\
$\Upsilon_{12}$	& 1	 	& 75		& 3 & 0 & 2 \\ 
\rule{0pt}{3ex}%
$\Omega_1$		& 1 		& 50 		& 4 & 0 & 2 \\
$\Omega_2$		& 1 		& \fiftybar 	  	& 3 & 0 & 0 \\
$\Omega_3$		& 1 		& 50 		& 1 & 0 & 2 \\
$\Omega_4$		& 1 		& \fiftybar 		& 8 & 0 & 0 \\
$\Pi_1$			& 1 		& 75 		& 6 & 0 & 2 \\
$\Pi_2$			& 1 		& 75 		& 3 & 0 & 0 \\
\hline
\end{tabular}
\caption{Superfields that break R symmetry and cause doublet-triplet splitting through the missing partner mechanism. }
\label{ta:RMPM}
\end{minipage}%
\qquad\begin{minipage}[b]{0.45\textwidth}
\captionsetup{width=\textwidth}
\centering
\begin{tabular}{| c | c c | c | c | c |}
\hline
\multirow{2}{*}{\rule{0pt}{4ex}Field}	& \multicolumn{5}{c |}{Representation} \\
\cline{2-6}
\rule{0pt}{3ex}			& $A_4$ 	& SU(5) & $\mathbb{Z}_9$ & $\mathbb{Z}_6$ & $\mathbb{Z}_4^R$ \\ [0.75ex]
\hline \hline
\rule{0pt}{3ex}%
$A_\mu$			& 3 		& 1 	& 3 & 0 & 2 \\
$A_\tau$			& 3 		& 1 	& 4 & 0 & 2 \\
$A_2$				& 3 		& 1 	& 7 & 0 & 2 \\
\rule{0pt}{3ex}%
$O_{e\mu}$			& 1 		& 1 	& 6 & 0 & 2 \\
$O_{e\tau}$			& 1 		& 1 	& 2 & 0 & 2 \\
$O_{\mu\tau}$			& 1 		& 1 	& 8 & 0 & 2 \\
$O_{e3}$			& 1 		& 1 	& 6 & 5 & 2 \\
$O_{23}$			& 1 		& 1 	& 5 & 2 & 2 \\
$O_{12}$			& 1 		& 1 	& 5 & 1 & 2 \\
$O_{13}$			& 1 		& 1 	& 3 & 3 & 2 \\
$O_{\mu5}$			& 1 		& 1 	& 0 & 4 & 2 \\
$O_{25}$			& 1 		& 1 	& 2 & 1 & 2 \\
$O_{\mu6}$			& 1 		& 1 	& 1 & 4 & 2 \\
$O_{56}$			& 1 		& 1 	& 7 & 2 & 2 \\
$O_{64}$			& 1 		& 1 	& 2 & 3 & 2 \\
$O_{14}$			& 1 		& 1 	& 4 & 3 & 2 \\
\rule{0pt}{3ex}%
$P_{ee}$ 		& 1 & 1 &  0 &  0 & 2 \\
$P_{\mu\mu}$ 		& 1 & 1 &  3 &  0 & 2 \\
$P_{22}$ 		& 1 & 1 &  7 &  0 & 2 \\
$P_{e4}$ 		& 1 & 1 &  7 &  5 & 2 \\
$P_{1e}$ 		& 1 & 1 &  6 &  4 & 2 \\
$P_{44}$ 		& 1 & 1 &  5 &  4 & 2 \\
$P_{34}$ 		& 1 & 1 &  4 &  4 & 2 \\
$P_{33}^{1,2}$ 		& 1 & 1 &  3 &  4 & 2 \\
$P_{2\tau}$ 		& 1 & 1 &  1 &  3 & 2 \\
\hline	
\end{tabular}
\caption{Driving superfields for the flavon alignment potential.\\}
\label{ta:driving}
\end{minipage}
\end{table}

\endgroup

Thus far we have assumed that the $A_4$ triplet VEVs are aligned in special directions. In this section we describe how these directions are obtained by the superpotential terms allowed by the symmetries.
In doing this, the role of $\mathbb{Z}_{6}$ becomes clearer. The driving sector, a set of superfields $ A_i $ and $ O_{ij} $ with $\mathbb{Z}_4^R$ charge 2, is listed fully in Table \ref{ta:driving}.%
\footnote{Note that the $O$ (and $P$) fields that carry no $\mathbb{Z}_{6}$ charge couple to $H_{5} H_{\bar{5}} \xi^n$ (with some power of $\xi$), e.g. $P_{22} H_{5} H_{\bar{5}}$. We do not discuss these further as the respective $F$-terms do not affect the alignment nor the origin of the $\mu$-term.}
The inclusion of the $ \mathbb{Z}_{6} $ symmetry is necessary because the driving superpotential, responsible for aligning the flavons, needs to have each driving field isolated,
as shown below (see also \cite{King:2013iva}):
\begin{align}
\begin{split}
W_{\mathrm{align}} 	&\sim A_\mu\phi_\mu\phi_\mu+A_\tau\phi_\tau\phi_\tau + A_2(\phi_2\phi_2+\phi_2\theta_1) \\ 
	&\qquad+ O_{e\mu}\phi_e\phi_\mu+O_{e\tau}\phi_e\phi_\tau+O_{\mu\tau}\phi_\mu\phi_\tau \\ 
	&\qquad+ O_{e3}\phi_e\phi_3+O_{23}\phi_2\phi_3+O_{12}\phi_1\phi_2+O_{13}\phi_1\phi_3 \\ 
	&\qquad+ O_{\mu 5}\phi_\mu\phi_5+O_{25}\phi_2\phi_5+O_{\mu 6}\phi_\mu\phi_6+O_{56}\phi_5\phi_6 \\ 
	&\qquad+ O_{64}\phi_6\phi_4+O_{14}\phi_1\phi_4.
	\label{flavon}
\end{split}
\end{align}
The additional $\mathbb{Z}_{6}$ charges ensure each term is separated from all others, leading to an array of vanishing $F$-term conditions that force mutual orthogonality conditions between many of the vacuum alignments. Since this was fully
discussed in \cite{King:2013iva}, we need only state the results here, namely that Eq.~\ref{flavon} leads to the following vacuum alignment patterns:
\begin{align}
\begin{array}{rlrlrl}
	\braket{\phi_e} & \sim \pmatr{1\\0\\0} 	\hspace{5ex} 	& \braket{\phi_\mu} 	& \sim \pmatr{0\\1\\0} 		\hspace{5ex} 	& \braket{\phi_\tau} 	& \sim \pmatr{0\\0\\1} \\ [5ex]
	\braket{\phi_1} & \sim \pmatr{2\\-1\\1} 			& \braket{\phi_2 }	& \sim \pmatr{1\\1\\-1} 			& \braket{\phi_3} 	& \sim \pmatr{0\\1\\1} \\ [5ex]
	\braket{\phi_4} & \sim \pmatr{1\\3\\1} 			& \braket{\phi_5}	& \sim \pmatr{1\\0\\1} 				& \braket{\phi_6 }	& \sim \pmatr{1\\0\\-1}.
\end{array}
\end{align}

The role of the VEVs (containing two zero entries) of the superfields $\phi_{e,\mu,\tau}$ appearing in Eq.~\ref{eq:dcl} was already discussed in Section \ref{sec:dcl}. Meanwhile, the role of the VEVs of the flavons  $\phi_{3,4}$ (redubbed $ \phi_{\mathrm{atm,sol}} $) was described in Section \ref{sec:neutrinos}. It is the special structure of these vacuum alignments,
combined with the phase of $ \physphase $ in the neutrino mass matrix, that leads to the very successful prediction of the leptonic mixing angles (as described in Section \ref{sec:fit}). The remaining VEVs are not directly relevant to the masses and mixings of SM fermions, but help shape the VEVs of $ \phi_{\mathrm{atm}} $ and $ \phi_{\mathrm{sol}} $.

\subsection{Flavon vacuum phases}
\label{sec:phasefixing}

With the direction of the $A_4$ triplet flavons $\phi$ fixed, we turn now to a discussion of how to fix the relative phase $\relphase \equiv \mathrm{arg}(v_{\mathrm{atm}} v_{\mathrm{sol}}^{\ast})$ to a discrete choice. We present a mechanism which does this by adding a number of fields $ P $ that are $ A_4 $ and $ SU(5) $ singlets, also given in Table \ref{ta:driving}, and which resemble the $ O $ fields except they do not force orthogonality between the flavons $ \phi $.  

These fields and their respective charge assignments result in the following invariant superpotential terms: 
\begin{align}
\begin{split}
W_{\mathrm{phase}} 	
	&\sim P_{ee}(\phi_e\phi_e+ M^2+P_{ee}^2)+P_{\mu\mu}(\phi_\mu\phi_\mu+Z_2 Z_3+P_{\mu\mu}^2) \\ 
	&\qquad+P_{e4}(\phi_e\phi_4+\theta_1\theta_2)+P_{22}(\phi_2\phi_2+\theta_1\theta_1) \\ 
	&\qquad+ P_{1e}(\phi_1\phi_e+\phi_6\phi_\tau)+P_{44}(\phi_4\phi_4+\phi_5\phi_\tau)+P_{34}(\phi_3\phi_4+\phi_6\phi_e)\\ 
	&\qquad+ P_{33}^{1,2}(\phi_3\phi_3+\phi_1\phi_\mu+\phi_5\phi_e)+ P_{2\tau}(\phi_2\phi_\tau+\phi_3\phi_6+\phi_4\phi_5),
	 \label{eq:fixing}
\end{split}
\end{align}
where each term technically has an associated real coupling $ \lambda $ which is $ \mathcal{O}(1) $ and may be made positive 
by field redefinitions. We omit these for simplicity as they have no effect on the general argument presented here, with one caveat: the two superfields $ P_{33}^{1,2}$ have exactly the same quantum numbers but different $ \lambda $ couplings to flavons. Due to this duplication there are two independent relations between the flavon VEVs
involving different $\lambda$ couplings which leads to an additional constraint on the phases of the respective VEVs. Exact values of these $ \lambda $ are not specified; it suffices that they are not equal.

Furthermore, the primary role of the SU(5) adjoint fields $ Z_2 $ and $ Z_3 $ which couple to $ P_{\mu\mu} $ is in the GUT breaking mechanism (as discussed in Section \ref{sec:GUTbreaking}). Their phases are fixed separately by other superpotential terms.

We begin the analysis of these terms by noting they do not affect the alignments of the flavons $ \phi$.
The corresponding $ F $-terms for each field $ P_{ij} $ produces a set of coupled equations that admit a solution where none of the  $A$, $O$, and $P$ fields but all the flavons obtain a VEV. Omitting the (real, positive, $\mathcal{O}(1)$)  $\lambda$ coefficients, these VEVs have the structure:
\begin{equation}
\begin{aligned}
v_e &\sim M & 
v_\mu &\sim (v_{Z_2} v_{Z_3})^\frac{1}{2} \\
v_\tau &\sim (v_{Z_2}~v_{Z_3})^{-\frac{1}{3}} M^\frac{5}{3} &
v_1 &\sim (v_{Z_2} v_{Z_3})^{-\frac{1}{2}}~v_3^2 \\
v_2 &\sim (v_{Z_2} v_{Z_3})^{-\frac{1}{6}} M^{-\frac{7}{3}}~v_3^3 \qquad&
v_4 &\sim (v_{Z_2} v_{Z_3})^{-\frac{1}{6}} M^\frac{1}{3}~v_3 \\
v_5 &\sim M^{-1}~v_3^2 &
v_6 &\sim (v_{Z_2} v_{Z_3})^{-\frac{1}{6}} M^{-\frac{2}{3}}~v_3^2 \\
v_{\theta_1} &\sim  (v_{Z_2} v_{Z_3})^{\frac{1}{6}}M^{-\frac{7}{3}}~v_3^3 &
v_{\theta_2} &\sim (v_{Z_2} v_{Z_3})^{-\frac{1}{6}} M^\frac{11}{3}~v_3^{-2} \\[1ex]
v_{O}&=~v_{P}=~v_A=~0.
\end{aligned}
\end{equation}
Regarding the magnitudes of the VEVs, two comments are in order. We assumed above that 
$M$  sets the scale of the VEV of $ \phi_e $, which is in contradiction with our previous assumption that 
it be $ \mathcal{O}(10^{-3}) M$. This violates our simplifying assumption that all mass scales are equal,
and demonstrates that some spectrum of mass scales is in fact required in this model.
As for the VEV $v_3$, as discussed earlier, it is driven to a specific scale $\Lambda_3$ radiatively \cite{Greene:1986jb}.
Writing $ \rho_i \equiv \mathrm{arg}(v_i)$, this VEV structure gives (up to multiples of $\pi$) the phase relation
\begin{equation}
\rho_4= \frac{2 \pi n}{3} - \frac{1}{6}(\rho_{Z_2} + \rho_{Z_3}) + \rho_3 ,
\label{fix}
\end{equation}
where $n$ is an integer, and similar relations for the other flavons as linear combinations of $ \rho_3 $, $(\rho_{Z_2} + \rho_{Z_3})$ and multiples of $ 2\pi/3 $. This is an important equation since it fixes the relative phase $\rho_3-\rho_4 = \relphase$ in terms of $\frac{1}{6}(\rho_{Z_{2}}+\rho_{Z_{3}})$. As discussed in Section \ref{sec:GUTbreaking}, $\rho_{Z_{2}}+\rho_{Z_{3}} =  \frac{2 \pi k'}{3}$, where we also establish that $\xiphase= \frac{2 \pi k}{9}$, for integers $k$, $k'$. From Eq.~\ref{eta}, $\physphase\equiv -\xiphase+2(\relphase )$, so we conclude that $\physphase$ is one of the nine complex roots of unity.

\section{GUT scale symmetry breaking, proton decay and the strong CP problem \label{sec:GUT}}

In this section we discuss the aspects of the model related to grand unification, starting with how the R-symmetry and the GUT gauge group are spontaneously broken. We refer to the superfields involved as the scalar sector; they are listed in Tables \ref{ta:SMF} and \ref{ta:RMPM}. We then describe the details of the MP mechanism, and finish this section with an analysis that justifies the absence of dangerous proton decay operators in the model.

\subsection{\secheadmath{SU(5)} and \secheadmath{\mathbb{Z}_4^R} breaking \label{sec:GUTbreaking}}
\begin{figure}[ht]
	\centering
	\begin{subfigure}{0.5\textwidth}
		\centering
		\includegraphics[scale=0.45]{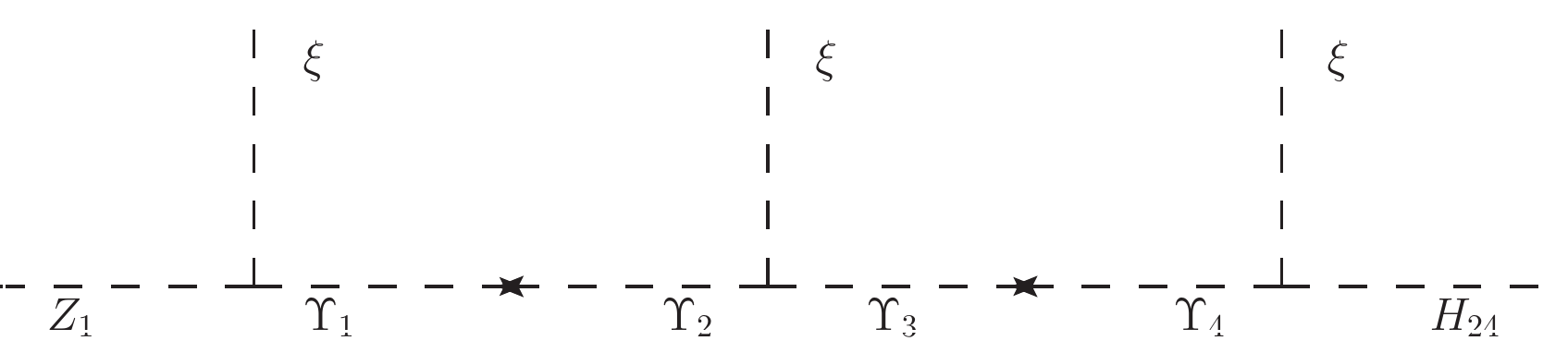}
		\caption{}
	\end{subfigure}%
	\begin{subfigure}{0.5\textwidth}
		\centering
		\includegraphics[scale=0.45]{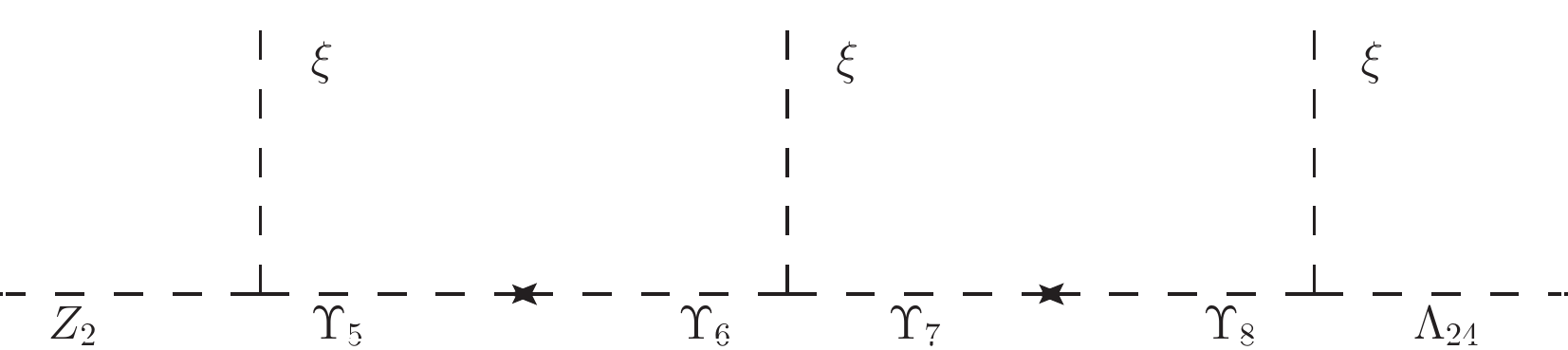}
		\caption{}
	\end{subfigure}
	\caption{Diagrams for non-renormalisable potential terms.}
	\label{fig:GUTpotcomplete}
\end{figure}

As previously discussed, the $\Upsilon$ messengers form pairs; their mass scale, unprotected by any symmetry, is near the highest scale of the theory, which we represent generically as $M$. The GUT breaking superpotential with non-renormalisable terms is then%
\footnote{A renormalisable term of the form $ Z_{2}H_{24}\Pi_2$, allowed by the symmetries, has been dropped to make the discussion more transparent. This term mixes the VEVs of the GUT breaking scalars with the ones in the MP mechanism so they should be naturally around the same scale ($M \sim\!M_{\mathrm{GUT}}$). Beyond this, its practical effect is minimal: the fields obtain VEVs with or without this term. Since the VEVs get very complicated when this ``mixing'' term is included, we ignore it for simplicity, simply bearing in mind that VEVs from both sets of fields are related.}%
\begin{align}
\label{eq:GUTpot}
\begin{split}
	W_{\mathrm{GUT}}	& = Z_1\left(M\Lambda_{24}+\frac{\lambda_1}{M^2}H_{24}\xi^3+\lambda_2 Z_1^2\right)+Z_2\left(\frac{\lambda_3}{M^2}\Lambda_{24}\xi^3+\lambda_4Z_2^2\right) \\
				& \qquad +Z_3\left(\lambda_5H_{24}^2+\lambda_6 Z_3^2\right).
\end{split}
\end{align}
We have five GUT adjoint superfields, three of which (the $Z$ fields) are charged by $2$ and two ($ \Lambda_{24} $ and $ H_{24} $) by $0$ under the R-symmetry. Also appearing in $W_{\mathrm{GUT}}$ is the Majoron $\xi$, the GUT singlet field which we have seen is involved in giving mass to several SM fermions and whose VEV breaks lepton number by giving the right-handed neutrinos their Majorana mass.
The supressions of the non-renormalisable terms in Eq.~\ref{eq:GUTpot} come precisely from the mass of the $\Upsilon$ messengers, as displayed in Fig.~\ref{fig:GUTpotcomplete}.

$W_{\mathrm{GUT}}$ has a non-trivial minimum:
\begin{equation}
\label{eq:GUTVEVs}
\begin{split}
	v_{Z_1} 	&=-\frac{i 2^{2/3} \lambda _1 \lambda _4^{1/3} \lambda _6^{1/6}}{3^{1/2} \lambda _2 \lambda _3 \lambda _5^{1/2}}M\\
	v_{Z_2} 	&=\frac{i 2^{1/3}  \lambda _1 \lambda _6^{1/6}}{3^{1/2} \lambda _2^{2/3} \lambda _3 \lambda _5^{1/2}}M \\
	v_{Z_3} 	&= \frac{i \lambda _1 \lambda _4^{1/3}}{3^{1/2} \lambda _2^{2/3} \lambda _3 \lambda _5^{1/2} \lambda _6^{1/6}}M
\end{split}\qquad
\begin{split}
	v_{H_{24}}	&=\frac{ \lambda _1 \lambda _4^{1/3} \lambda _6^{1/3}}{\lambda _2^{2/3} \lambda _3 \lambda _5}M \\
	v_{\Lambda_{24}} &=\frac{2^{1/3} \lambda _1^2 \lambda _4^{2/3} \lambda _6^{1/3}}{\lambda _2 \lambda _3^2 \lambda _5} M\\
	\langle \xi\rangle^3 	&=\frac{2^{1/3} \lambda _4^{1/3}}{\lambda _2^{1/3} \lambda _3}M^3,
\end{split}
\end{equation}
where all the adjoint scalars get a VEV of the form $\braket{\Phi_{24}}=v_{\Phi_{24}}~\mathrm{diag}(2,2,2,-3,-3)$.
By themselves, the $F$-terms associated with $W_{\mathrm{GUT}}$ also allow a trivial minimum where the magnitude of each VEV vanishes. But after SUSY is broken and we consider the effects of the small contribution from radiative breaking \cite{Ibanez:1982fr} to the scalar components of the GUT breaking superfields (as we did in Section \ref{sec:A4} for the $A_4$ breaking flavons), the stationary point with vanishing magnitudes is no longer a minimum due to the radiatively induced negative squared mass term. To a very good approximation the true minima are given by the magnitudes in Eq.~\ref{eq:GUTVEVs}, which are now a lower energy state than the trivial $ F $-term solution.

We conclude that Eq.~\ref{eq:GUTpot} can generate GUT and R-symmetry breaking at high scale, 
with $\mathbb{Z}_4^R$ broken to $\mathbb{Z}_2^R$ (standard R-parity preserved) by the $Z_i$ VEVs. Because  $\mathbb{Z}_4^R$  is broken at a high scale, the no-go theorem from \cite{Fallbacher:2011xg} does not apply to our model and we verified that all the components of the $SU(5)$ adjoints acquire GUT scale masses. 

A slightly unappealing issue with $W_{\mathrm{GUT}}$ 
is that the minimum requires some non-${\cal O}(1)$ choice of $\lambda$ parameters if we are to obtain a hierarchy between the VEVs of $H_{24}$ and $\Lambda_{24}$, and an appropriate value for $ \braket{\xi}\!/M $ as shown in Eq.~\ref{eq:vevfit}. These requirements come from the successful fit to up and down quark and charged lepton masses, as discussed in Sections \ref{sec:u}, \ref{sec:dcl} and \ref{sec:fit}, and partly also for the $\mu$ term, as will be discussed shortly.
However, since the messengers will in general have different masses (recall we set them all equal to $M$ only for simplicity), the $\lambda$ parameters need not be as hierarchical as Eq.~\ref{eq:GUTVEVs} appears to indicate. For example, if the masses of messengers $\Sigma$ are slighly larger than the GUT scale masses of messengers $\Upsilon$, this would allow all $\lambda$ to be $\mathcal{O}(1)$.

We note also that, although we are considering a situation where the superpotential parameters (the $M$ and $\lambda$ couplings) are real due to CP conservation, the VEVs of the GUT breaking scalars may be complex, since they depend on $n^\mathrm{th}$ order roots of real numbers. As has been shown in Sections \ref{sec:neutrinos} and \ref{sec:phasefixing}, the phases of the fields $ \xi $, $ Z_2 $ and $ Z_3 $ are relevant for establishing the physical phase $ \physphase $ in the neutrino mass matrix, which controls neutrino masses and mixing. We see immediately that $ \xiphase = \frac{2\pi k}{9} $, for integer $ k $, i.e. one of nine roots of unity. While $ \rho_{Z_2} $ and $ \rho_{Z_3} $ individually can be any of six roots, originating in the factor $ \lambda_5^{1/6} $, their product $ Z_2 Z_3 $ cancels this factor such that the largest root is a third, giving $ \rho_{Z_2} + \rho_{Z_3} = \frac{2\pi k'}{3} $, for integer $ k' $. 

\subsection{Doublet-triplet splitting, Higgs mixing and the \secheadmath{\mu} term \label{sec:splitting}}

Given that we have a number of GUT representations containing (SM gauge group) $SU(2)$ doublets and triplets we turn now to a brief discussion of how doublet-triplet splitting is achieved in this model. Although one could alternatively introduce the double MP mechanism \cite{Hisano:1994fn} as demonstrated in \cite{Antusch:2014poa},
here we limit ourselves to describing how the MP mechanism is implemented with the fields listed in Table \ref{ta:RMPM}.

\begin{figure}[!ht]
\centering
\includegraphics[scale=.6]{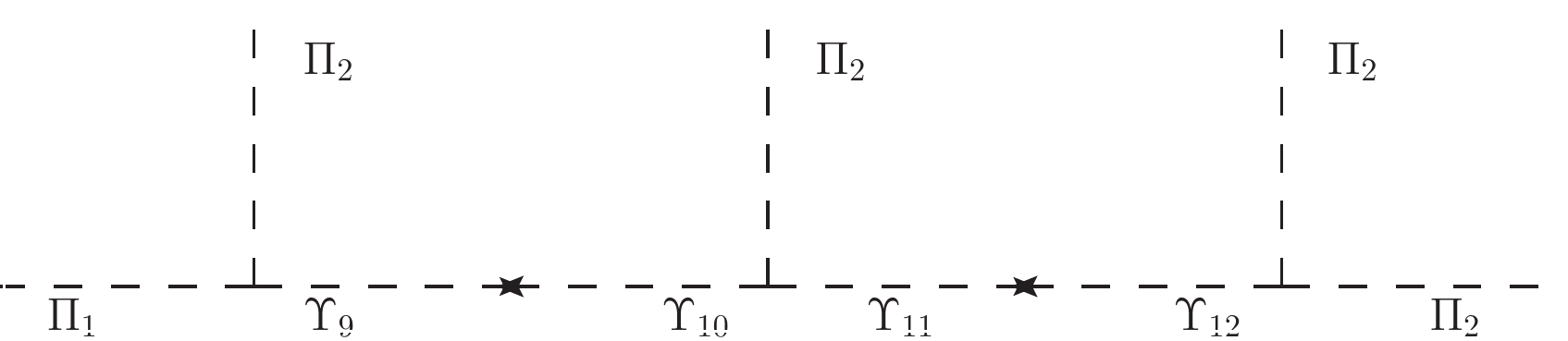}   
\caption{Diagram for non-renormalisable term in the 75 potential.}
\label{fig:W75}
\end{figure}

We have a superpotential
\begin{equation}
	W_\Pi = \Pi_1\left(\lambda_7\Pi_1^2+ M\Pi_2+\frac{\lambda_8}{M^2}\Pi_2^4\right),
\label{eq:W75}
\end{equation}
which gives $\Pi$, the \textbf{75}s of $SU(5)$, their VEVs
\begin{equation}
\label{eq:75VEV}
v_{\Pi_1}	=-\frac{ 1}{ {16}^{1/3} \lambda _7^{1/2} \lambda_8^{1/6}}~M, \qquad
v_{\Pi_2} 	=-\frac{ 1}{4 \lambda_8^{1/3}}M,
\end{equation}
which are aligned with the SM singlet inside the $SU(5)~\textbf{75}$. The non-renormalisable term in $W_\Pi$ comes from the diagram in Fig.~\ref{fig:W75}. 

With Eq.~\ref{eq:75VEV}, the MP mechanism proceeds through the superpotential:
\begin{align}
\begin{split}
	W_{\mathrm{MP}} 	&\sim H_{\bar 5} \Omega_1\Pi_2+H_5\Omega_2\Pi_1+\xi\Omega_1\Omega_2 \\ 
			&\qquad + H_{\bar{45}}\Omega_3\Pi_2+M H_{\bar{45}}H_{45}+M\Omega_3\Omega_4 \\
			&\qquad + H_{\bar 5} H_{45}\Pi_2+H_5 H_{\bar{45}}\Pi_1\left(\frac{\xi}{M}\right)^8.
\end{split}
\end{align}
The very high order non-renormalisable term at the end arises through the $\Sigma$ messengers already listed in Table \ref{ta:Mess} (half of which participate in the up quark Yukawa terms, as discussed in Section \ref{sec:u} and illustrated in Fig.~\ref{fig:upmass}).
Strictly speaking this term does not participate in splitting the masses of $SU(2)$ doublets and triplets, rather it is the source of the $\mu$ term in our model, as shown below. 

The terms in $W_{\mathrm{MP}}$ generate mixing between the \textbf{45}s and \textbf{5}s of $SU(5)$.
The mass matrix for the $SU(2)$ triplets contained in the \textbf{45}s and \textbf{5}s is:
\begin{align}
\begin{split}
	M_{\mathbf{3}} 		&= \overline{\textbf{3}}^T \pmatr{0 &v_{\Pi_2} &v_{\Pi_2} & 0 \\ v_{\Pi_1}\tilde{\xi}^8 & M & 0 & v_{\Pi_2} \\ v_{\Pi_1} & 0 & \braket{\xi} &0 \\0 &0 & 0 & M} \textbf{3} \\[2ex] 
	\overline{\textbf{3}}^T 	&=\pmatr{\textbf{3}(H_{\bar 5})&\textbf{3}(H_{\bar{45}})& \textbf{3}(\Omega_2)& \textbf{3}(\Omega_4)} \\
	\textbf{3}^T 		&=\pmatr{\textbf{3}(H_5) &\textbf{3}(H_{45}) & \textbf{3}(\Omega_1) & \textbf{3}(\Omega_3)},
\end{split}
\end{align}
where once again $ \tilde{\xi} = \xivev/M $. Taking $\braket{\Pi_{1,2}}\sim M$, the eigenvalues of this mass matrix are all of order $M$ (i.e. at the GUT scale), leading us to conclude there are no light $SU(2)$ triplets. Conversely, for the doublets we have the matrix:
\begin{align}
	\pmatr{\mathbf{2}(H_{\bar 5}) & \mathbf{2}(H_{\overbar{45}})} \pmatr{0 & v_{\Pi_2} \\ v_{\Pi_1} \tilde{\xi}^8 & M} \pmatr{\mathbf{2}(H_{5}) \\ \mathbf{2}(H_{45})}.
\end{align}
It is clear that were it not for $\tilde{\xi}^8$, the determinant of this mass matrix would vanish. We may rotate to the basis of the MSSM Higgs doublets $H_{u,d}$ and a pair of very heavy doublets $H_{u,d}^H$:
\begin{align}
\begin{split}
	\pmatr{\mathbf{2}(H_{\bar{5}}) \\ \mathbf{2}(H_{\overbar{45}})} &\approx \frac{1}{\sqrt{2}}\pmatr{1 & -1 \\ 1  & 1} \pmatr{H_d^H \\ H_d} \\
	\pmatr{\mathbf{2}(H_{5}) \\ \mathbf{2}(H_{45})} &\approx \pmatr{\tilde{\xi}^8 & 1 \\ -1  & \tilde{\xi}^8} \pmatr{H_u^H \\ H_u}.
\end{split}
\end{align}

The usual MSSM term $\mu H_d H_u$ comes from this mechanism with:
\begin{equation}
\mu\sim\frac{v_{\Pi_1}v_{\Pi_2}}{M}\tilde\xi^8,
\end{equation}
where $v_{\Pi_1}$ provides the necessary $\mathbb{Z}_4^R$ breaking. Using the fit from Eq.~\ref{eq:vevfit} we see that $\tilde{\xi}^8 \sim 1.6\times 10^{-10} M_{\mathrm{GUT}}$. If we choose the couplings at the vertices of the tower that generates the $\xi^8$ term to be $\sim 0.5$ we may get a term $\mu\sim \mathcal{O}(10^2-10^3)$ GeV without any fine-tuning.

\subsection{Proton decay}

A classic problem in GUT theories, and in particular those based on $SU(5)$, is the prediction of excessively fast proton decay. The most dangerous processes come from the ``dimension 5'' operators $TTTF$ (for a discussion of dimension 6 operators we refer the reader to \cite{Antusch:2014poa}).
The ``dimension 5'' operators are forbidden by the symmetries of the model, but related higher order
operators of the following form are allowed: 
\begin{equation}
 T_i T_j T_k F \frac{Z \phi}{M^3}\left(\frac{\xi}{M}\right)^{n_{ijk}},
\label{proton}
\end{equation}
where the extra superfields shown are needed for such terms to be invariant under the symmetries. Since we are working with the renormalisable theory, in order for this type of effective term to be present at the GUT scale, there must be messengers allowing them. In this case, an analysis of the $SU(5)$ index structure revels there should either be messengers that are $SU(5)$ $\overbar{\textbf{10}}$, or $SU(5)$  $\textbf{5}$ that are also charged under $\mathbb{Z}_4^R$. As one can confirm from Table \ref{ta:Mess}, our model has neither: $\overbar{\textbf{10}}$ messengers were not used, and the $\textbf{5}$ messengers are all neutral under $\mathbb{Z}_4^R$. We conclude therefore that our symmetry content, together with the existing set of messengers, do not allow any such GUT scale suppressed operators that would lead to excessively fast proton decay to be generated. The operators in Eq.~\ref{proton} may in principle be generated by physics at the Planck scale, with the scale $M$ replaced by the Planck mass, leading to highly suppressed proton decay.

\subsection{Strong CP problem and the Nelson-Barr resolution}

We first recall the strong CP problem, namely that the physical angle $\overline{\theta}=\theta_{\mathrm{QCD}}-\theta_q$, where $\theta_{\mathrm{QCD}}$ multiplies the topological gluon term $(g_s^2/32\pi^2)G\tilde{G}$ and $\theta_q= \arg \det (Y^uY^d)$, is limited to be $\overline{\theta} <10^{-10}$ by the non-observation of the neutron EDM \cite{Beringer:1900zz, Baker:1997ed}. The origin of such a small number, $\overline{\theta} <10^{-10}$, is commonly called the strong CP problem. It is interesting to compare this with the CP violation related to the weak interaction in the quark sector; the relevant quantity is the Jarlskog invariant $J^q \sim  \det[Y^u Y^{u\dagger},Y^dY^{d\dagger}]$, which, when compared to data, is required to be non-vanishing, and indeed in the standard parameterisation, requires a large phase angle $\delta^q \sim 1$.

It turns out that our model resolves the strong CP problem without relying on the introduction of axions. Unlike the axion solution, which requires a global $U(1)$ symmetry with a colour anomaly, we shall rely on the fact that the high energy theory conserves CP, ensuring that $\theta_{\mathrm{QCD}}=0$. CP is then spontaneously broken in such a way as to yield $\delta^q\sim 1$ while maintaining $\overline{\theta} <10^{-10}$ and in particular $ \theta_q <10^{-10}$. How it achieves this feat can be seen from Eq.~\ref{eq:yufull} where $Y^u$ is real, while the structure of $Y^d$ in Eq.~\ref{eq:ydfull} gives it a real determinant. This is due to the lack of a Yukawa term $ Y^d_{21} $, meaning the coupling $ Y^d_{12} $ (which is the only complex Yukawa coupling) does not appear in the determinant of $ Y^d $. Therefore there are no contributions to $ \theta_q$ even after spontaneous CP breaking. This is similar to the Nelson-Barr mechanism \cite{Nelson:1983zb, Barr:1984qx}, where the triangular form of Yukawa matrices was proposed, although in our model $\theta_q$ vanishes due to the triangular form of $Y^d$ only, with $Y^u$ being non-triangular and real.

For a successful resolution of the strong CP problem, we must ensure that no higher order corrections to the Yukawa matrices arise which would violate the bound $ \theta_q <10^{-10}$. The main focus is on the Yukawa coupling $Y^d_{21}$ which is zero at leading order but which may in principle receive higher order corrections, violating the triangular structure. However in our model such higher order corrections are absent at the field theory level with the specified messenger sector. This entry in the Yukawa matrix would arise from the coupling of the bilinears $T_2 H_{\bar 5, \overbar{45}}$ to the bilinear $\phi_e F$. Since these terms are non-renormalisable, we require messengers to form them. The messengers that could produce such terms are the $ X_i $ fields in Table \ref{ta:Mess}. With these messengers, the only allowed connection to $\phi_e F$ is $T_1 H_{\bar 5}$ (contributing to $ Y^d_{11} $), thus forbidding the $ Y^d_{21} $ term, even when allowing for all higher-order corrections. Therefore the specified model has no strong CP violation arising from $Y^d_{21}$ since the required operators are not generated at the field theory level.

It is also important to consider the effect of higher order corrections arising from the Planck scale, since such operators only have to respect the symmetries of the model, and do not require the specified messenger sector to generate them. The biggest contribution would come from the term%
\footnote{This term would also give a contribution to lepton angles of $\mathcal{O}(10^{-3})$ which is negligible.}
\begin{equation}
	T_2 H_{\bar 5}\frac{\phi_e}{M_P}F.
\end{equation}
With a general choice of phase, such a term would lead to $\theta_q\sim 10^{-4}$ which is far too big. However the contribution to $\theta_q$ may be avoided by a judicious choice of GUT breaking phases. As stated in section \ref{sec:fit}, the physical phase in the down quark Yukawa matrix is $\zeta=\rho_\xi-2\rho_{H_{24}}-\rho_{\Lambda_{24}}$. The new Planck suppressed term has a  phase $\zeta'=-\rho_\xi+2\rho_{\Lambda_{24}}$. Choosing a relation between phases $2\rho_{H_{24}}=\rho_{\Lambda_{24}}$, then $\zeta=-\zeta'$ and this way the contribution to $\theta_{q}$ vanishes. This happens for one in three cases. The next biggest contribution comes from a term
\begin{equation}
	T_2 H_{\bar{45}}\frac{\xi^2\phi_e}{M_P^3}F,
\end{equation}
giving $\theta_q\sim 10^{-14}$ which is several orders of magnitude below the current experimental bound. Any other Planck suppressed terms allowed by the symmetries are further suppressed so we need not consider them. Therefore the model may resolve the strong CP problem even in the presence of Planck scale operators controlled only by symmetry. Finally, extra contributions may come from SUSY breaking terms. If we assume that there is no extra CP violation in this sector, which is controlled by the spontaneously CP violating flavons, the SUSY flavour problem is under control and such contributions to $\overline{\theta}$ are also expected to be negligible \cite{Antusch:2013rla}.

\section{The leptogenesis link}
\label{sec:link}

The link between leptogenesis and the PMNS matrix was first studied for sequential dominance in \cite{King:2002qh}. In the original form of CSD, the columns of the Dirac mass matrix in the flavour basis were orthogonal to each other and consequently the CP asymmetries for cosmological leptogenesis \cite{DiBari:2012fz} vanished \cite{Antusch:2006cw,King:2006hn,Bertuzzo:2009im, Choubey:2010vs}. In this model, leptogenesis does not vanish since the columns of the Dirac mass matrix in the flavour basis are not orthogonal. 

Interestingly, since the seesaw mechanism in this model with two right-handed neutrinos only involves a single phase $\physphase=2\pi /3$, both the leptogenesis asymmetries and the neutrino oscillation phase must necessarily originate from this phase, providing a direct link between the two CP violating phenomena in this model.

Following the arguments in \cite{Antusch:2006cw}, the produced baryon asymmetry $Y_B$ from leptogenesis in the seesaw model in Eq.~\ref{seesaw} satisfies
\begin{equation}\label{eq:BAU}
	Y_B \propto \pm\sin \physphase,
\end{equation}
where the ``$+$'' sign applies to the case $M_{\rm atm}  \ll M_{\rm sol}$ and the ``$-$'' sign holds for the case $M_{\rm sol} \ll M_{\rm atm}$. Since the observed baryon asymmetry $Y_B$ is positive, it follows that, for $M_{\rm atm}  \ll M_{\rm sol}$, we must have $\sin \physphase$ to be positive, while for $M_{\rm sol} \ll M_{\rm atm}$ we must have $\sin \physphase$ to be negative. We have seen that positive $\physphase$ is associated with negative $\delta^l$ and {\it vice versa}. Although the global fits do not distinguish the sign of $\physphase$, the present hint that $\delta^l \sim -\pi/2$ would require positive $\physphase$, then in order to achieve positive $Y_B$ we require $M_{\rm atm}  \ll M_{\rm sol}$, which is natural in our model, corresponding to ``light sequential dominance'', where successful leptogenesis may be achieved in the two right-handed neutrino model as discussed in \cite{Antusch:2011nz}.

\section{Conclusion \label{sec:con}}

We have presented here a fairly complete realisation of an $SU(5)$ GUT flavoured with 
$A_4$, which leads to the MSSM plus two right-handed neutrinos below the GUT scale.
The $A_4$ family symmetry unifies the three families of 5-plets $F$ and its vacuum alignment 
determines the Yukawa matrices.
In addition a $\mathbb{Z}_9\times \mathbb{Z}_{6}$ symmetry provides the mass hierarchy and controls 
spontaneous CP violation in both the quark and lepton sectors while a $\mathbb{Z}_4^R$ symmetry 
is broken to $\mathbb{Z}_2^R$, identified as the usual R-parity.
Proton decay is under control in this model, with the symmetries forbidding dangerous dimension-5 operators, and similar (but higher order) operators being very suppressed.
The strong CP problem is resolved in a similar way to the Nelson-Barr mechanism.
The model is highly predictive and satisfies many distinct (and non-trivial) phenomenological requirements.

Imposing CP at the high scale is an important feature of the model. If we do not impose CP then
all couplings become complex, leading to all VEVs having undetermined phases.
In particular the phase $ \eta $, present in both neutrino mixing and leptogenesis, is no longer restricted to a discrete choice. However the link between leptogenesis and low energy phenomenology remains. 
On the other hand we would no longer solve the strong CP problem.

We highlight the ubiquitous nature of the flavon field $\xi$ across all the sectors of the model: it triggers
spontaneous CP violation in both the quark and lepton sectors, generates up-type quark mass hierarchies and CKM mixing, explains the smallness of down quark and electron masses and breaks lepton number, providing 
the hierarchy between solar and atmospheric right-handed neutrino masses.
In addition, the $\xi$ field is responsible for generating the small (complex) $\mu$ term. 
The phase of the $\xi$ field VEV also contributes to the relative phase $\physphase$
appearing in the neutrino mass matrix. 
This phase, the only one
appearing in the neutrino mass matrix and in the formula
for the baryon asymmetry of the universe,
provides a direct link between the PMNS phase $\delta^l$ and leptogenesis.

We emphasise that the entire PMNS matrix is predicted without any free parameters, up to nine choices for $ \physphase $, where we select $ \physphase = +2\pi/3$, since it is preferred by comparing CSD3 to data.
The required vacuum alignments are provided from $A_4$.
The model effectively serves to 
yield the CSD3 scheme (with two right-handed neutrinos) within a fully working and viable SUSY GUT of flavour
in which all quark and lepton masses and mixings are successfully described. 
Indeed the model provides an excellent fit (better than one sigma) to all quark and lepton (including neutrino) masses, mixing and CP violation. All fermion mass hierarchies are understood in the sense that purportedly
$\mathcal{O}(1)$ couplings indeed contain no strong hierarchies. However the most immediate 
predictions of the model are those of CSD3 with two right-handed neutrinos
and $ \physphase = +2\pi/3$, 
namely a normal neutrino mass hierarchy with $m_1=0$, a reactor angle of $\theta^l_{13}\approx 8.7^{\circ}$,
a solar angle $\theta^l_{12}\approx 34^{\circ}$, close to maximal
atmospheric mixing $\theta^l_{23}\approx 46^{\circ}$ and almost maximal leptonic CP violation, with 
an oscillation phase $\delta^l\approx -87^{\circ}$ consistent with the current 
hint $\delta^l\approx -\pi/2$.

The reason why the field content is so large is that the model is fairly complete.
In particular it is {\em renormalisable at the GUT scale}, which requires a large explicit field content including many heavy messenger superfields. It also addresses many aspects relevant both to a GUT and to family symmetry models (stopping short of discussing the details of SUSY breaking and its string theory completion).
In particular, the $A_4$, $SU(5)$ and $R$-symmetry symmetry breaking sectors all require large field content. 
For example, the GUT symmetry is 
broken by an explicit superpotential at the GUT scale, including doublet-triplet splitting via
a missing partner mechanism (leaving no light exotic degrees of freedom at the low scale), 
Higgs mixing and the origin of the MSSM $\mu$ term of the right order of magnitude.

Despite the many successes of the model, there are inevitably several important issues
that lie beyond the scope of this paper. 
To take one example, we do not discuss GUT scale threshold corrections, which will
be important in maintaining successful gauge coupling unification in the presence of many fields, including
colour triplets, at the GUT scale. 
In fact all the additional superfields in non-trivial representations of 
the gauge group may have masses at or above the GUT scale.
Another important issue is that of the low energy superpartner spectrum in this model. While we expect 
SUSY induced flavour changing to be under control for the 5-plets, which are unified into an $A_4$
triplet, this is not the case for the 10-plets $T_i$ which are singlets of $A_4$,
leading to flavour violation in the super-CKM basis.
It would be interesting to study this in the future.

\section*{Acknowledgements}

This project has received funding from the European Union's Seventh Framework Programme for research, technological development and demonstration under grant agreement no PIEF-GA-2012-327195 SIFT.
The authors also acknowledge partial support from the European Union FP7 ITN-INVISIBLES (Marie Curie Actions, PITN- GA-2011- 289442) and CONACyT.

\appendix
\section{\secheadmath{A_4} basis convention}
\label{A4}

In the basis we are using (see \cite{Ma:2001dn} for more details), one has the following Clebsch-Gordan rules for the multiplication of two triplets, 
$3\times 3 = 1+1'+1''+3_1+3_2$, 
\begin{equation}\label{pr}
\begin{array}{lll}
(ab)_1&=&a_1b_1+a_2b_2+a_3b_3 \\
(ab)_{1'}&=&a_1b_1+\omega a_2b_2+\omega^2a_3b_3 \\
(ab)_{1''}&=&a_1b_1+\omega^2 a_2b_2+\omega a_3b_3 \\
(ab)_{3_1}&=&(a_2b_3,a_3b_1,a_1b_2) \\
(ab)_{3_2}&=&(a_3b_2,a_1b_3,a_2b_1)\,
\end{array}
\end{equation}
where the components of the two triplets are
given by  $a=(a_1,a_2,a_3)$ and $b=(b_1,b_2,b_3)$, 
and $\omega^3=1$.


\end{document}